\documentclass[11pt]{article}

\include{setup}

\allowdisplaybreaks[1]

\begin{document}

\thispagestyle{empty}
\def\thefootnote{\fnsymbol{footnote}}
\setcounter{footnote}{1}
\null
\draftdate\hfill MPP-2006-149 \\
\strut\hfill PITHA 06/11 \\
\strut\hfill hep-ph/0611353
\vskip 0cm
\vfill
\begin{center}
  {\Large \boldmath{\bf
      MSSM Higgs-boson production in bottom-quark fusion:\\[.5em]
      electroweak radiative corrections }
\par} \vskip 2.5em
{\large
{\sc Stefan Dittmaier}\\[1ex]
{\normalsize \it Max-Planck-Institut f\"ur Physik (Werner-Heisenberg-Institut), \\
D-80805 M\"unchen, Germany}\\[2ex]
{\sc Michael Kr\"amer, Alexander M\"uck}\\[1ex]
{\normalsize \it Institut f\"ur Theoretische Physik E, RWTH Aachen, \\ 
D-52056 Aachen, Germany}\\[2ex]
{\sc Tobias Schl\"uter}\\[1ex]
{\normalsize \it Max-Planck-Institut f\"ur Physik (Werner-Heisenberg-Institut), \\
D-80805 M\"unchen, Germany}}\\[2ex]
\par \vskip 1em
\end{center}\par

\vskip .0cm \vfill {\bf Abstract:} \par Higgs-boson production in
association with bottom quarks is an important discovery channel for
supersymmetric Higgs particles at hadron colliders for large values
of $\tan\beta$. We present the complete ${\cal O}(\alpha)$
electroweak and ${\cal O}(\alpha_{\mathrm{s}})$ strong corrections
to associated bottom--Higgs production through $\Pb\bar \Pb$ fusion
in the MSSM and improve this next-to-leading-order prediction by
known two-loop contributions to the Higgs self-energies, as provided
by the program {\tt FeynHiggs}.  Choosing proper renormalization and
input-parameter schemes, the bulk of the corrections (in particular
the leading terms for large $\tan\beta)$ can be absorbed into an
improved Born approximation.  The remaining non-universal
corrections are typically of the order of a few per cent. Numerical
results are discussed for the benchmark scenarios SPS~1b and SPS~4.
\par
\vskip 1cm
\noindent
March 2007
\par
\null
\setcounter{page}{0}
\clearpage
\def\thefootnote{\arabic{footnote}}
\setcounter{footnote}{0}

\section{Introduction}
The Higgs mechanism is a cornerstone of the Standard Model (SM) and
its supersymmetric extensions. The masses of the fundamental
particles, electroweak gauge bosons, leptons, and quarks, are
generated by interactions with Higgs fields. The search for Higgs
bosons is thus one of the most important endeavours in high-energy
physics and is being pursued at the upgraded proton--antiproton
collider Tevatron with a centre-of-mass (CM) energy of $1.96$~TeV,
followed in the near future by the proton--proton collider LHC with
$14$~TeV CM energy.

Various channels can be exploited to search for Higgs bosons at hadron
colliders. Higgs radiation off bottom quarks~\cite{Raitio:1978pt}
\begin{equation}
\Pp\bar{\Pp} / \Pp\Pp \to \Pb\bar{\Pb}\,\PHgeneric \!+\!X
\label{eq:procs_hadron}
\end{equation}
is the dominant Higgs-boson production mechanism in
su\-per\-sym\-me\-tric theories at large $\tan\beta$, where the
bottom--Higgs Yukawa couplings are, in general, strongly enhanced.
With $\PHgeneric = \PH$, $\Phn$, $\PHn$, $\PAn$ we denote the SM
Higgs boson or any of the neutral Higgs bosons of the Minimal
Su\-per\-sym\-me\-tric Standard Model (MSSM).  Current searches for
MSSM bottom--Higgs associated production at the Fermilab Tevatron
widely exclude values $\tan\beta \gsim 50$ for light $\MA \approx
100\GeV$~\cite{Abulencia:2005kq,Carena:2005ek}, depending in detail
on the value of the Higgs mixing parameter $\mu$.

Two different formalisms have been employed to calculate the cross
section for associated $\Pb\bar{\Pb}\,\PHgeneric$ production.  In a
four-flavour number scheme (4FNS) with no $\Pb$ quarks in the
initial state, the lowest-order QCD production processes are
gluon--gluon fusion and quark--antiquark annihilation, $\Pg\Pg \to
\Pb\bar \Pb\,\PHgeneric$ and $\Pq\bar \Pq \to \Pb\bar
\Pb\,\PHgeneric$, respectively. The inclusive cross section for
$\Pg\Pg \to \Pb\bar \Pb\,\PHgeneric$ develops large logarithms
$\sim\ln(\mu_F/m_\Pb)$, which arise from the splitting of gluons
into nearly collinear $\Pb\bar \Pb$ pairs. The large scale $\mu_F
\sim \MHgeneric$ corresponds to the upper limit of the collinear
region up to which factorization is valid. The $\ln(\mu_F/m_\Pb)$
terms can be summed to all orders in perturbation theory by
introducing bottom parton densities. This defines the so-called
five-flavour number scheme (5FNS)~\cite{Barnett:1987jw}. The use of
bottom distribution functions is based on the approximation that the
outgoing $\Pb$ quarks are at small transverse momentum. In this
scheme, the leading-order (LO) process for the inclusive $\Pb \bar
\Pb\,\PHgeneric$ cross section is $\Pb\bar \Pb$ fusion,
\begin{equation}
\Pb\bar \Pb\to \PHgeneric\,.
\label{eq:procs_bbfus}
\end{equation}
The next-to-leading order (NLO) cross section in the 5FNS includes
${\cal O}(\alpha_{\mathrm{s}})$ corrections to $\Pb\bar \Pb\to
\PHgeneric$ and the tree-level processes $\Pg\Pb\to\Pb\PHgeneric$ and
$\Pg\bar\Pb\to\bar\Pb\PHgeneric$.  

To all orders in perturbation theory the four- and five-flavour
schemes are identical, but the way of ordering the perturbative
expansion is different, and the results do not match exactly at finite
order. However, numerical comparisons between calculations of
inclusive Higgs production in the two schemes
\cite{Dittmaier:2003ej,Campbell:2004pu,Dawson:2005vi,Buttar:2006zd}
show that the two approaches agree within their respective
uncertainties, once higher-order QCD corrections are taken into
account.

There has been considerable progress recently in improving the
cross-section predictions for inclusive associated $\Pb\bar
\Pb\,\PHgeneric$ production by calculating
NLO-QCD~\cite{Dittmaier:2003ej,Dawson:2005vi} and
SUSY-QCD~\cite{Hollik:2006vn} corrections in the four-flavour scheme,
and NNLO QCD
corrections~\cite{Dicus:1998hs,Harlander:2003ai} in the five-flavour
scheme. The inclusion of higher-order effects is crucial for an
accurate theoretical prediction and, eventually, a determination of
Higgs-boson parameters from the comparison of theory and experiment.
In this paper we further improve the cross-section prediction and
present the first calculation of the complete ${\cal O}(\alpha)$
electroweak corrections to associated bottom--Higgs production through
$\Pb\bar \Pb\to \PHgeneric$ in the MSSM.

The complete one-loop QCD and electroweak corrections for the decay
of MSSM Higgs bosons to bottom quarks have been presented in
Ref.~\cite{Dabelstein:1995js} more than a decade ago. While the
predictions for Higgs-boson decays have been improved and refined in
recent years, supersymmetric QCD and electroweak corrections to the
production cross section have so far only been investigated at the
level of universal corrections for large $\tan\beta$ (see
e.g.~\citeres{Carena:2005ek,Hahn:2006my}).

In Section~\ref{se:tree} we shall set the notation for the
supersymmetric model. The calculation of the NLO QCD and electroweak
corrections is described in some detail in Sections~\ref{se:nlo},
\ref{se:ewcorr}, and \ref{se:ewmssmcorr}. Numerical results for MSSM
Higgs-boson production at the LHC are presented in
Section~\ref{se:pheno}.  We conclude in Section~\ref{se:conclusion}.
The Appendices provide details on the scenarios of the
supersymmetric model under consideration and present further
numerical results.

\section{Radiative corrections}
\label{se:corrections}

\subsection{Tree-level Yukawa couplings and cross section}
\label{se:tree}

In the Standard Model, Higgs production from bottom-quark fusion is
governed by the interaction term
\begin{equation}
\la_{\bar{\Pb} \Pb \PH } = - \lambda^{\rm SM}_\Pb \, \, 
\bar{\Pb} \, \Pb \, \PH \, , 
\end{equation}
where $\lambda_\Pb$ is the bottom-quark Yukawa coupling and $\PH$
denotes the field of the physical Higgs boson. The corresponding
mass term for the bottom quark is generated by the Higgs vacuum
expectation value $\vev$, leading to the tree-level relation
\begin{equation}
\label{eq:smyukawa}
\lambda^{\rm SM}_\Pb = \frac{\Mb}{\vev} = \, \frac{e \, \Mb}{2 \sw \MW} , 
\end{equation}
where $e$ is the electromagnetic coupling constant, $\sw$ the sine
of the weak mixing angle, and $\MW$ the W-boson mass.

For the MSSM we will follow the conventions of
Ref.~\cite{Haber:1984rc}, where the two Higgs doublets are denoted
as 
\begin{equation}
\PHd = \left( \begin{array}{c}
              \mathswitchr{h^+_d} \\
              \frac{1}{\sqrt{2}} (\vevd + \mathswitchr{h^0_d} + \ri \mathswitchr{\chi^0_d})\\
              \end{array}
              \right),
\qquad
\PHu = \left( \begin{array}{c}
              \frac{1}{\sqrt{2}} (\vevu + \mathswitchr{h^0_u} - \ri \mathswitchr{\chi^0_u})\\
              -\mathswitchr{h^-_u} \\
              \end{array}
              \right).
\end{equation} 
The bottom quarks couple to $\PHd$, giving mass to the down-type
quarks via the vacuum expectation value $\vevd$. Masses for up-type
quarks are generated by a second Higgs doublet $\PHu$ with vacuum
expectation value $\vevu$.  Considering the MSSM without
CP-violating phases, the CP-even neutral Higgs-boson fields $\Phn$
and $\PHn$ are linear combinations of $\mathswitchr{h^0_d}$ and
$\mathswitchr{h^0_u}$.  One conventionally defines the Higgs mixing
angle $\alpha$ by writing
\begin{equation}
\label{eq:mixiningmatrix}
\Vector{\Phn \\ \PHn} = \Matrix{\cosa & - \sina \\
                              \sina &  \cosa} 
                              \Vector{\mathswitchr{h^0_u} \\ 
\mathswitchr{h^0_d}} \, .
\end{equation}
Here and in the following, we will frequently use the notation
$\sina\equiv \sin \alpha$, $\cosa\equiv \cos \alpha$ and
generalizations thereof. The vacuum expectation values are
parameterized according to $\vevu = \vev \, \sin \beta$, $\vevd =
\vev \, \cos \beta$, i.e.
\begin{equation}
\tanb \equiv \tan \beta \equiv \frac{\vevu}{\vevd} \quad \mr{and} \quad 
\vev^2 \equiv \vevu^2 + \vevd^2 \, .
\end{equation}
Taking the pseudoscalar Higgs mass $\MA$ and $\tanb$ as input
parameters of the MSSM Higgs sector, one finds the tree-level
relation
\begin{equation}
\label{eq:sa_tree_level}
t_{2 \alpha} = \frac{\MA^2+\MZ^2}{\MA^2-\MZ^2} \, \, t_{2 \beta} \, \, \, 
\end{equation}
with $s_{2 \alpha} < 0$.  The fields of the physical neutral
pseudoscalar Higgs boson $\PAn$ and the neutral would-be Goldstone
boson $\mathswitchr{G^0}$ are given by
\begin{equation}
\Vector{\PAn \\ \mathswitchr{G^0}} = \Matrix{\cosb & -\sinb \\
                                             \sinb &  \cosb} 
\Vector{\mathswitchr{\chi^0_u} \\ \mathswitchr{\chi^0_d}} \, .
\end{equation}
Consequently, the $\Pb \bar{\Pb} \Phn$, $\Pb \bar{\Pb} \PHn$, and
$\Pb \bar{\Pb} \PAn$ couplings read
\begin{equation}
\label{eq:mssmyukawa}
\begin{array}{ccccc}
\lambda_\Pb^{\Phn} & = & \frac{\displaystyle - \sina \,\Mb}{\displaystyle \vevd} 
                   & = & - \lambda^{\rm SM}_\Pb \,\frac{\displaystyle \sina}{\displaystyle \cosb} \, , \\[1mm]
\lambda_\Pb^{\PHn} & = & \frac{\displaystyle \cosa \,  \Mb}{\displaystyle \vevd} 
                   & = & \lambda^{\rm SM}_\Pb \,
                         \frac{\displaystyle \cosa}{\displaystyle \cosb} \, , \\[1mm]
\lambda_\Pb^{\PAn} & = & \frac{\displaystyle - \sinb \, \Mb}{\displaystyle \vevd} 
                   & = & - \lambda^{\rm SM}_\Pb \, \tanb \, .
\end{array}
\end{equation} 
Hence, the Yukawa couplings are enhanced for large values of
$\tanb$. Note that for large masses of the pseudoscalar Higgs boson,
$\Phn$ is known to be SM like and $\sina \to -\cosb$.

The leading-order partonic cross sections are given by
\begin{equation}
\label{eq:tree}
\hat{\sigma}^{0}_{\PHgeneric} 
\, = 
\, \frac{\pi}{6} \, \frac{(\lambda^{\rm SM}_\Pb)^2}{\MHgeneric^2} \, \, \delta(1-\tau) \, 
           \left\{ \begin{array}{l} \sina^2/\cosb^2 \\ 
                                    \cosa^2/\cosb^2 \\
                                    \tanb^2 \end{array} 
\quad \mathrm{for} \, \, \, \, \begin{array}{l} \Phn \\ \PHn \\ \PAn \end{array} \, \, 
\mathrm{production},
           \right. 
\end{equation}
where $\PHgeneric =(\Phn,\PHn,\PAn)$, $\tau = \MHgeneric^2/\hat{s}$,
$\sqrt{\hat{s}}$ is the partonic CM energy, and the incoming bottom
quarks are treated as massless particles in accordance with QCD
factorization.

\subsection{SUSY-QCD corrections}
\label{se:nlo}
The QCD corrections to the process $\Pb \bar{\Pb}\to \Phn,\PHn,\PAn$
are known to NNLO~\cite{Dicus:1998hs,Harlander:2003ai}, with a small
residual QCD factorization and renormalization scale uncertainty of
less than $\sim 10$\%. If one chooses the renormalization and
factorization scales as $\mu_R=\MH$ and $\mu_F=\MH/4$, respectively,
the impact of the genuine NNLO QCD corrections is typically less
than 5\%~\cite{Harlander:2003ai}.  We have reproduced the NLO QCD
result and extend previous analyses by including the ${\cal
O}(\alpha_s)$ SUSY-QCD corrections from virtual squark and gluino
exchange.

The $\overline{\mathrm{MS}}$ scheme has been adopted for the
renormalization of the bottom-quark mass $\Mb$ and for the
factorization of initial-state collinear singularities. The
renormalization of the bottom--Higgs Yukawa coupling is fixed in
terms of the bottom-mass renormalization.  In order to sum large
logarithmic corrections $\propto \ln(m_\Pb/\mu_R)$ we evaluate the
Yukawa coupling with the running $\Pb$-quark mass
$\overline{m}_{\Pb}(\mu_R)$~\cite{Braaten:1980yq}.

The ${\cal O}(\alpha_s)$ SUSY-QCD corrections comprise self-energy
and vertex diagrams induced by virtual sbottom and gluino exchange,
as shown in Fig.~\ref{NLOdiagramsSUSY}.
\begin{figure}
  \parbox{8cm}{\input{graphs_SE_SUSY.tex}}
  \parbox{8cm}{\input{graphs_SUSY.tex}}
  \vspace*{-3em}
  \mycaption{\label{NLOdiagramsSUSY} Self-energy (a) and vertex (b)
    corrections from gluino exchange. Sbottom mass eigenstates are
    denoted as $\tilde{\Pb}_i$ with $i=1,2$.}
\end{figure}
It is well
known~\cite{Hall:1993gn,Hempfling:1993kv,Carena:1994bv,Pierce:1996zz}
that the SUSY-QCD corrections are enhanced for large $\tanb$.  This
important effect can be qualitatively understood as follows.  
Unlike down-type quarks, which only couple to the down-type Higgs
field at tree level, the down-type squarks also couple to the
up-type Higgs field via terms in the superpotential. The
corresponding coupling strength is proportional to the enhanced
Yukawa coupling $\Mb/\vevd$ times the Higgs mixing parameter $\mu$.
Hence, the vacuum expectation value $\vevu$ of $\PHu$ leads to
a mixing term in the sbottom mass matrix proportional to
$\mu\,\tanb$, which dominates the sbottom mixing
angle~$\Theta_{\tilde{\Pb}}$,
\begin{equation}
\label{eq:sbottom_mixing_angle}
\sin (2 \Theta_{\tilde{\Pb}}) = 
\frac{2 \, \Mb \, \left( A_\Pb - \mu \, t_{\beta} \right)}
{m^2_{\tilde{\Pb}_1} - m^2_{\tilde{\Pb}_2}} \, ,
\end{equation}
where $A_\Pb$ denotes the soft-supersymmetry-breaking trilinear
scalar coupling and $\tilde{\Pb}_{1,2}$ are the sbottom mass
eigenstates.  The factor $\sin (2 \Theta_{\tilde{\Pb}})$ directly
enters the (scalar part of the) $\Pb$-quark self-energy, which in
turn  enters the Yukawa coupling renormalization via the $\Pb$-mass
counterterm $\de\Mb$.  The corresponding mass shift is usually
denoted by $-\Mb \Delta_\Pb$ with
\begin{equation}
\label{eq:qcd_tanb_enhanced}
\Delta_\Pb = \frac{C_F}{2}\frac{\alpha_s}{\pi} \, 
m_{\tilde{g}} \, \mu \, t_{\beta} \, 
I(m_{\tilde{\Pb}_1},m_{\tilde{\Pb}_2},m_{\tilde{g}}) \, ,
\end{equation}
$C_F = 4/3$, and the auxiliary function
\begin{equation}
\label{eq:I}
I(a,b,c) = \frac{-1}{(a^2-b^2)(b^2-c^2)(c^2-a^2)} \left(
a^2 b^2 \ln \frac{a^2}{b^2} + 
b^2 c^2 \ln \frac{b^2}{c^2} + 
c^2 a^2 \ln \frac{c^2}{a^2}
\right) \, .
\end{equation}
Here, $m_{\tilde{g}}$ is the gluino mass.  As shown in
\citere{Carena:1999py} by power counting in
$\alpha_{\mathrm{s}}\,\tanb$, the contribution of $\Delta_\Pb$ can
be summed by the replacement
\begin{equation}
\label{eq:replacement}
\Mb \to \frac{\Mb}{1 + \Delta_\Pb}
\end{equation}
in the bottom Yukawa coupling.  As explained above, the loop-induced
coupling of the up-type Higgs to bottom quarks also involves a
factor $\alpha_{\mathrm{s}}\,\tanb$.  The full contribution to the
$\PHgeneric \Pb\bar\Pb$ vertex receives an additional factor
$\{c_\alpha, s_\alpha, c_\beta\}$ from the $\PHu$ part in
$\PHgeneric =\{\Phn,\PHn,\PAn\}$.  Power counting in
$\alpha_{\mathrm{s}}\,\tanb$ shows~\cite{Carena:1999py} that the
$\tanb$-enhanced vertex corrections of the form $(\alpha_s \,
\tanb)^n$ are one-loop exact, i.e.\ they do not appear at higher
orders ($n \ge 2$).  Collecting all $\tanb$-enhanced corrections to
the bottom Yukawa couplings leads to the effective couplings
$\overline{\lambda}{}_\Pb^{\PHgeneric}$~\cite{Carena:1999py,Guasch:2003cv},
where
\begin{eqnarray}
\label{eq:Yukawa_shift}
\frac{\overline{\lambda}{}_\Pb^{\Phn}}{\lambda_\Pb^{\mr{SM}}} & = & 
- \frac{\sina}{\cosb} \,
\frac{1 - \Delta_\Pb/(\tanb \tana)}{1 + \Delta_\Pb} \, ,
\nonumber\\
\frac{\overline{\lambda}{}_\Pb^{\PHn}}{\lambda_\Pb^{\mr{SM}}} & = &  
\frac{\cosa}{\cosb} \, 
\frac{1 + \Delta_\Pb \, \tana/\tanb}{1 + \Delta_\Pb} \, ,
\\
\frac{\overline{\lambda}{}_\Pb^{\PAn}}{\lambda_\Pb^{\mr{SM}}} & = &  
- \tanb \,
\frac{1 - \Delta_\Pb/\tanb^2}{1 + \Delta_\Pb}
  \, . \nonumber
\end{eqnarray} 
Note that $\overline{\lambda}{}_\Pb^{\Phn}$ is still SM like for
large $\MA$, independent of the large-$\tanb$ summation owing to
$\tanb\tana\to-1$ in this limit. The summation formalism can be
extended~\cite{Guasch:2003cv} to include corrections proportional to
the trilinear coupling $A_\Pb$ in (\ref{eq:sbottom_mixing_angle}).
However, these corrections turn out to be small and summation
effects may thus safely be neglected for the MSSM scenarios under
consideration in this work.

To combine the features of the above effective treatment with the
complete one-loop SUSY-QCD calculation, we modify the
renormalization scheme to absorb the above corrections into a
redefinition of the bottom mass in the Yukawa coupling. Hence, an
additional counterterm
\begin{equation}
\frac{\bar{\delta}{m_{\Pb}^\Phn}}{\Mb} = \Delta_\Pb \, \left(1+\frac{1}{\tana\tanb}\right) \, , \, \quad 
\frac{\bar{\delta}{m_{\Pb}^\PHn}}{\Mb} = \Delta_\Pb \, \left(1-\frac{\tana}{\tanb}\right) \, , \, \quad 
\frac{\bar{\delta}{m_{\Pb}^\PAn}}{\Mb} = \Delta_\Pb \, \left(1+\frac{1}{\tanb^2}\right)
\end{equation}
is added for $\Phn$, $\PHn$, and $\PAn$ production, respectively, to
remove the $\tanb$-enhanced contributions from the explicit one-loop
result in order to avoid double-counting. We use the convention of
\citere{Dabelstein:1995js} where $m_{\Pb,0}=\Mb+\de\Mb$.

As we shall demonstrate in the numerical analysis presented in
Section~\ref{se:pheno}, the SUSY-QCD radiative corrections are
indeed sizeable at large $\tanb$.  After summation of the
$\tanb$-enhanced terms, however, the remaining one-loop SUSY-QCD
corrections are negligibly small, at the level of per mille and
below.

\subsection{Electroweak SM corrections and calculational details}
\label{se:ewcorr}
The electroweak corrections naturally decompose into a purely
photonic, QED-like part $\delta_{\mathrm{QED}}$ and the remaining
weak contributions $\delta_{\mathrm{weak}}$: $\delta_{\mathrm{ew}} =
\delta_{\mathrm{QED}}+\delta_{\mathrm{weak}}$.  Each of these
contributions forms a gauge-invariant subset of the ${\cal
O}(\alpha)$-corrected cross section. The photonic corrections due to
virtual photon exchange and real photon emission can be obtained
from the QCD results by appropriately adjusting colour factors and
electric charges. For the QED renormalization of the bottom mass we
use the on-shell scheme, because electroweak running effects beyond
the one-loop level are negligible.

The divergences due to collinear photon emission from the massless
$\Pb$-quarks are removed by mass factorization as in QCD, i.e.\ by a
redefinition of the bottom parton densities according to
\begin{equation}
\begin{split}
  f_{\Pb}(x) \to f_{\Pb}(x,\mu_F) & + \int_{x}^{1} \, \frac{dz}{z} \, \,
  f_\Pb \left(\frac{x}{z},\mu_F \right) \, Q_{\Pb}^2 \, \frac{\alpha}{2
    \pi} \, \left\{ \left[ P_{qq}(z) \right]_+
    \,  \left( \Delta + \ln \frac{\mu^2}{\mu_F^2} \right) \, - C_{qq}(z) \right\} \, \\
  & + \int_{x}^{1} \, \frac{dz}{z} \,  f_\gamma \left(\frac{x}{z},\mu_F
  \right) \, 3 \, Q_{\Pb}^2 \, \frac{\alpha}{2 \pi} \, \left\{
    P_{q\gamma}(z) \,  \left( \Delta + \ln \frac{\mu^2}{\mu_F^2} \right) \,  
    - C_{q\gamma}(z) \right\} \, ,
\end{split}
\end{equation}
where $\Delta=1/\epsilon - \gamma_E + \ln(4 \pi)$ is the standard
divergence in $D=4-2\epsilon$ dimensions, $\gamma_E$ is Euler's
constant, $Q_{\Pb} = -1/3$ is the electric $\Pb$ charge, and $\mu_F$
denotes the QED factorization scale which is identified with the QCD
factorization scale but is chosen independently of the scale $\mu$
introduced by dimensional regularization. The factor 3 in the second
line stems from the splitting of the photon into $\Pb \Pbbar$-pairs
of different colour. Furthermore, 
\begin{equation}
P_{qq}(z) = \frac{1+z^2}{1-z}  \quad \mathrm{and} \quad
P_{q\gamma}(z) =  z^2 + (1-z)^2 
\end{equation}
are the quark and photon splitting functions, respectively, and
$C_{qq}$, $C_{q\gamma}$ the coefficient functions specifying the
factorization scheme. Following standard QCD terminology one
distinguishes \MSbar\ and DIS schemes defined by
\begin{eqnarray}
C^{\overline{\mathrm{MS}}}_{qq}(z) & = & C^{\overline{\mathrm{MS}}}_{q\gamma}(z)  = 0,
\nonumber \\
C^{\rm DIS}_{qq}(z) & = & \left[ P_{qq}(z) \left( \ln \frac{1-z}{z} -
                   \frac{3}{4} \right) + \frac{9+5 z}{4} \right]_+ \, \, ,\\
C^{\rm DIS}_{q\gamma}(z) & = & P_{q\gamma}(z) \ln \frac{1-z}{z} -
                     8 z^2 + 8 z -1 \, \, . \nonumber
\end{eqnarray}
The equivalent factorization procedure using fermion masses as
regulators instead of dimensional regularization can, e.g., be found
in~\citere{Diener:2005me}. In our numerical analysis we employ the
MRST2004qed parton distribution functions~\cite{Martin:2004dh},
which include ${\cal O}(\alpha)$ corrections defined in the DIS
factorization scheme~\cite{Diener:2005me}. The MRST2004qed
pa\-ra\-me\-teri\-za\-tion also provides a photon density necessary
to compute the hadronic cross section for the ${\cal O}(\alpha)$
photon-induced processes $\gamma \, \Pb \to \PHgeneric \, \Pb$ and
$\gamma \, \Pbbar \to \PHgeneric \, \Pbbar$.

We have calculated the NLO QCD and QED corrections using dimensional
regularization and alternatively using a mass regularization for the
collinear divergences, where we employed the methods of
\citeres{Diener:2005me,Dittmaier:1999mb} for the mass regulators.

Similar to the QCD case, the QED corrections are universal for the
production of SM and MSSM Higgs bosons. Their size is quite small
since the potentially large correction due to collinear photon
emission is absorbed into the parton distribution function
$f_{\Pb}(x,\mu_F)$.  We shall discuss numerical results in
Section~\ref{se:pheno}.

Let us now turn to the remaining electroweak corrections
$\delta_\mathrm{weak}$. As we shall discuss in more detail in
Section~\ref{se:MSSM_results}, the small but finite bottom mass can
induce sizeable corrections in the electroweak sector of the MSSM due
to additional ($\tanb$-enhanced) bottom Yukawa couplings in loops.
Hence, while we neglect the $\Pb$-mass at tree level and for the QCD
and QED corrections as required by QCD factorization, we keep the
finite bottom mass $\Mb$ in the calculation of the relative one-loop
weak correction $\delta_\mathrm{weak}$. Thus, our result for
$\delta_\mathrm{weak}$ also contains kinematical $\Mb$ effects that
formally lie beyond the accuracy of the 5FNS-calculation.  However,
these effects are small. Different choices for the numerical value
of the bottom mass used within the calculation of the relative
one-loop correction lead to results which formally differ by NNLO
effects.  Because the $\Mb$-dependence is dominated by the strength
of the Yukawa coupling, we have chosen the running bottom mass (as
defined after summation of $\tanb$-enhanced terms) as input.

Both in the SM and in the MSSM, the one-loop electroweak corrections
have been calculated independently using two different approaches:
In one approach, the Feynman-diagrammatic expressions for all
self-energies and one-loop vertex diagrams have been generated using
the program package {\tt FeynArts}~\cite{Hahn:2000kx}. The
calculations have then been performed with the help of the program
package {\tt FormCalc}~\cite{Hahn:1998yk}, and the loop integrals
have been evaluated numerically with  {\tt
LoopTools}~\cite{Hahn:1998yk}.  In a second approach, also starting
from the amplitudes generated by {\tt FeynArts}, all calculations,
including the evaluation of the loop integrals, have been performed
with a completely independent set of in-house routines.  The two
calculations are in mutual agreement.  We note that the
regularization of the complete electroweak MSSM corrections has been
performed using both constrained differential renormalization as
implemented in  {\tt FormCalc} as well as dimensional reduction. 
Both regularization procedures are known to be equivalent at the
one-loop level~\cite{Hahn:1998yk}, and the results of the two
calculations are in agreement. We refrain from displaying the
complete analytic results and restrict ourselves to a discussion of
the renormalization conditions and the input-parameter schemes.

Using standard notation, the vertex counterterm at one-loop order 
is given by 
\begin{equation}
\label{eq:EW_CT_generic}
\delta^{\Pb \Pbbar \PH}_{\mathrm{CT}} 
\, = \, \delta Z_e + \frac{\delta Z_H}{2} + 
\frac{\delta \Mb}{\Mb} + \frac{\delta Z_L^\Pb + \delta Z_R^\Pb}{2}
- \frac{\delta \sw}{\sw} - \frac{\delta \MW^2}{2 \MW^2} \, .
\end{equation}
Employing the on-shell renormalization scheme, this results in 
\begin{equation}
\label{eq:EW_CT}
\begin{split}
\delta^{\Pb \Pbbar \PH}_{\mathrm{CT}} 
= & \, \delta Z_e + \frac{\delta Z_H}{2} + 
\Sigma^{\Pb}_S (\Mb^2) - 2 \Mb^2 \left( \Sigma^{\prime \, \Pb}_S (\Mb^2) + 
                                      \Sigma^{\prime \, \Pb}_V (\Mb^2) \right) \\[2ex]
&                                     
+ \frac{1}{2} \left( 
\frac{\cw^2 - \sw^2}{\sw^2} \frac{\Sigma_{\PW}(\MW^2)}{\MW^2}
- \frac{\cw^2}{\sw^2} \frac{\Sigma_{\PZ}(\MZ^2)}{\MZ^2}
\right)  \, ,
\end{split}
\end{equation} 
where $\Sigma^{\Pb}_{S,V}$ denotes the scalar and vector part of the
$\Pb$-quark self-energy, respectively, and $\Sigma^{\prime}$ refers
to the derivative of the self-energy with respect to the external
momentum squared.  The relation $\sw^2 = 1 - \MW^2/\MZ^2$ is used to
determine $\delta \sw$. Here and in Section~\ref{se:ewmssmcorr} we
only consider the real part of the self-energies and follow the
conventions of Ref.~\cite{Dabelstein:1994hb}. For compactness, the
transverse parts of gauge-boson self-energies  are simply written as
$\Sigma_{\PW}$, $\Sigma_{\PZ}$, etc.

The different input-parameter schemes are specified by the choice of
$\delta Z_e$. In the $\alpha(0)$-scheme, using the low-energy 
fine-structure constant $\alpha(0)$ as input,  we
have~\cite{Denner:1993kt}\footnote{We follow the convention  of
Haber and Kane~\cite{Haber:1984rc}  to define the covariant
derivative for SU(2)$_\mathrm{L}$, i.e.\ the sign in front of
$\Sigma_{\gamma Z}$ in  (\ref{eq:delta_Z_e_alpha_0}) differs
from~\citere{Denner:1993kt}.}
\begin{equation}
\label{eq:delta_Z_e_alpha_0}
\left. \delta Z_e \right|_{\alpha(0)} = 
             \frac{1}{2}\Sigma^{\prime}_{\gamma} (0) + 
             \frac{\sw}{\cw} \frac{\Sigma_{\gamma Z}(\MZ^2)}{\MZ^2} \, .
\end{equation} 
In the $\alpha(0)$-scheme, $\delta Z_e$ contains logarithms of the
light fermion masses inducing large corrections of the form $\alpha
\ln (\Mf^2/\hat{s})$, which are related to the running of the
electromagnetic coupling $\alpha(Q)$ from $Q=0$ to a high energy
scale. 
In order to correctly reproduce the non-perturbative hadronic part
in this running, which enters via $\Pi_{\gamma}(\MZ^2)-\Pi_{\gamma}(0)$
with $\Pi_{\gamma}(k^2) = \Sigma_{\gamma}(k^2)/k^2$
denoting the photonic vacuum polarization,
we adjust the quark masses to the asymptotic tail of 
$\Pi_{\gamma}(k^2)$.  
In the $\alpha(\MZ)$-scheme, using $\alpha(\MZ)$ as defined
in~\citere{Burkhardt:1995tt} as input, this adjustment is implicitly
incorporated, and the counterterm reads
\begin{equation}
\left. \delta Z_e \right|_{\alpha(\MZ)} = 
\left. \delta Z_e \right|_{\alpha(0)} - \frac{1}{2} \Delta \alpha (\MZ^2) \, ,
\end{equation}
where
\begin{equation}
\Delta \alpha (Q^2) = 
\Pi^{f \ne t}_{\gamma}(0) - \Re \Pi^{f \ne t}_{\gamma}(Q^2) \, ,
\end{equation}       
and $\Pi^{f \ne t}_{\gamma}$ denotes the photonic vacuum
polarization induced by all fermions other than the top quark.  
Hence, in the $\alpha(\MZ)$-scheme, the final result does not depend
on the above logarithms of the light fermion masses.  In the
$G_\mu$-scheme, $\alpha$ is determined from the muon-decay constant
$\GF$ according to
\begin{equation}
\alpha_{G_\mu}=\frac{\sqrt{2} G_\mu \MW^2 s_\PW^2}{\pi} = 
\alpha(0) (1 + \Delta r) .
\label{eq:alphagmu}
\end{equation}
The radiative QED corrections for muon decay in the framework of the
effective 4-fermion interaction are already encoded in the numerical
value for $G_\mu$. The additional corrections from a full one-loop
SM calculation are taken into account through $\Delta
r$~\cite{Sirlin:1980nh} according to
\begin{equation}
\label{eq:Delta_r}
\left. \delta Z_e \right|_{G_\mu} = 
\left. \delta Z_e \right|_{\alpha(0)} - \frac{1}{2} \Delta r \, .
\end{equation}
Since $\Delta \alpha (\MZ^2)$ is explicitly contained in $\Delta r$,
the large fermion-mass logarithms are also absent in the
$G_\mu$-scheme.  
Moreover, since the lowest-order cross section is proportional to 
$\alpha/s_\PW^2$ for the production of all the three Higgs bosons, 
in $G_\mu$-parameterization it absorbs the large universal correction 
$\Delta\rho$ from the $\rho$-parameter, which is $\propto\GF\Mt^2$ 
and represents a part of $\Delta r$. Dividing Eq.~\refeq{eq:alphagmu}
by $s_\PW^2$, one easily sees that $\alpha_{G_\mu}/s_\PW^2$
absorbs $\Delta r$ and thus also $\Delta\rho$. 
We will use the $G_\mu$-scheme unless stated
otherwise.  We have also performed two independent calculations for
$\Delta r$ in the MSSM, using either constrained differential
renormalization or dimensional reduction, and find agreement with
the  result of \citere{Chankowski:1993eu}.

In the SM, the Higgs mass can be defined by an on-shell
renormalization condition. The wave-function renormalization of the
Higgs-boson field is conveniently chosen in the on-shell scheme,
\begin{equation}
\label{eq:HigssCT}
\delta Z_{\PH} 
\, = \, - \, \Sigma^{\prime}_{\mathrm{H}} (\MH^2) \, ,
\end{equation}
where $\Sigma_{\mathrm{H}}$ is the Higgs-boson self-energy.

\subsection{Electroweak MSSM corrections}
\label{se:ewmssmcorr}

The photonic corrections in the MSSM do not change with respect to
the SM case. For the conventions and for the renormalization of the
MSSM Higgs sector, which is more involved, we essentially follow
\citere{Dabelstein:1994hb}.%
\footnote{For clarity, we specify our conventions for some
  field-theoretic quantities where the conventions of
  \citere{Dabelstein:1994hb} might be unclear. Explicit tadpole
  vertex functions for Higgs fields are denoted as $\Ga^{\phi^0}=\ri
  T^{\phi^0}$, i.e.\ $T^{\phi^0}$ differs from
  \citere{Dabelstein:1994hb} by a global sign. The $\PAn\PZ$ mixing
  self-energy is derived from the vertex function
  $\Ga^{\PAn\PZ}_\mu(k,-k) = k_\mu \Sigma_{\PAn\PZ}(k^2)$, where $k$
  is the incoming $\PAn$ momentum.}  In particular, a proper
renormalization scheme has to be specified to determine the vertex
counterterm $\delta^{\Pb \Pbbar \PH}_{\mathrm{CT}}$, cf.\ 
(\ref{eq:EW_CT_generic}), including the renormalization of $\tanb$.
The wave-function renormalization for the Higgs doublet fields is
usually defined by
\begin{equation}
\PH_i \to Z_{\PH_i}^{1/2} \PH_i = \PH_i \left( 1 \, + \, 
\frac{1}{2} \, \delta Z_{\PH_i} \right)
\end{equation}
with $i= \Pu,\Pd$. For the vacuum expectation values of the Higgs
fields one defines
\begin{equation}
\vev_i \to Z_{\PH_i}^{1/2} (\vev_i - \delta \vev_i)  = \vev_i \left( 1 \, + \, 
\frac{1}{2} \, \delta Z_{\PH_i}  - \frac{\delta \vev_i}{\vev_i} \right) \, ,
\end{equation}
where the last equation is valid to one-loop order with $Z_{\PH_i} =
1 + \delta Z_{\PH_i}$.  The freedom of wave-function renormalization
can then be used to impose the condition
\begin{equation}
\frac{\delta \vevu}{\vevu} \, = \, \frac{\delta \vevd}{\vevd} \, 
\end{equation}
leading to 
\begin{equation}
\label{eq:tanb_ren}
\frac{\delta \tanb}{\tanb} = 
\frac{1}{2} \left( \delta Z_{\PHu} - \delta Z_{\PHd} \right) \, .
\end{equation}
Hence, for the MSSM Higgs sector the counterterm $\delta Z_\PH$
in~(\ref{eq:EW_CT_generic}) reads
\begin{equation}
\delta Z_\PH \, = \, \frac{1}{2} \, \delta Z_{\PHd} 
+ \sinb^2 \, \frac{\delta \tanb}{\tanb} = \frac{1}{2} 
\left( \cosb^2 \,\delta Z_{\PHd} + \sinb^2 \,\delta Z_{\PHu}  \right) \, .
\end{equation}
Note, that $\delta Z_\PH$ includes all parts of the vertex
counterterm that are related to the Higgs sector, i.e.\ $\delta
\tanb$ as well as the wave-function renormalization counterterm
$\delta Z_{\PHd}$. The quantity $\delta Z_\PH$ should not be
confused with the wave-function renormalization constants for the
physical Higgs fields.  Since the counterterm $\delta Z_\PH$ is
universal for the production of $\Phn$, $\PHn$, and $\PAn$ via
down-type quarks, the whole vertex counterterm $\delta^{\Pb \Pbbar
\PH}_{\mathrm{CT}}$ (\ref{eq:EW_CT_generic}) is also universal.

\bigskip

We consider two renormalization schemes for $\tanb$:
\begin{itemize}
\item[i)] Following Dabelstein~\cite{Dabelstein:1994hb} and
  Chankowski et al.~\cite{Chankowski:1992er}, a vanishing on-shell
  $\PAn \PZ$-mixing can be used as a renormalization condition, i.e.
\begin{equation}
\hat{\Sigma}_{\PAn \PZ} (\MA^2) \, = \, 0 \, ,
\end{equation}
where 
\begin{equation}
\hat{\Sigma}_{\PAn \PZ} (k^2) \,=\,
\Sigma_{\PAn \PZ} (k^2) - \MZ s_{2\beta} \frac{\delta\tanb}{\tanb}
\end{equation}
denotes the real part of a renormalized self-energy defined
according to the conventions in~\citere{Dabelstein:1994hb}. To fix
the second wave-function renormalization constant, one demands the
on-shell condition
\begin{equation}
\label{eq:wf_ren_A0}
\hat{\Sigma}^{\prime}_{\PAn} (\MA^2) \, = \, 0 
\end{equation}
for the residue of the $\PAn$-boson propagator.  From the last two
equations, one finds 
\begin{equation}
\begin{split}
\label{eq:fieldstrengthren}
\delta Z_{\PHu} & = \, - \Sigma^{\prime}_{\PAn} (\MA^2) 
+ \frac{\tanb}{\MZ} \, \Sigma_{\PAn \PZ} (\MA^2) \, ,\\
\delta Z_{\PHd} & = \, -\Sigma^{\prime}_{\PAn} (\MA^2) -
\frac{1}{\tanb\MZ} \, \Sigma_{\PAn \PZ} (\MA^2) \, ,
\end{split}
\end{equation} 
which defines the DCPR
scheme~\cite{Dabelstein:1994hb,Chankowski:1992er} for the $\tanb$
renormalization [see (\ref{eq:tanb_ren})]. \item[ii)] Alternatively,
in the \DRbar\ scheme, the counterterm for $\tanb$ is proportional
to $\Delta=1/\epsilon - \gamma_E + \ln(4 \pi)$.  Hence, one can
convert (\ref{eq:fieldstrengthren}) to the \DRbar\ scheme by setting
the remaining finite pieces of $\Sigma_{\PAn \PZ}$ to zero. 
Accordingly, $\tanb^{\overline{\mathrm{DR}}}$ is a
renormalization-scale-dependent  quantity, i.e.\ the input for a
given model has to be fixed at a given scale $\mu_R$. To one-loop
order, the conversion of the $\tanb$ input parameters from the
\DRbar\ scheme to  $\tanb^{\, \mathrm{DCPR}}$ in the DCPR scheme is
given by
\begin{equation}
\label{eq:TB_conversion}
\biggl[ \tanb + \frac{1}{2 \MZ c_{\beta}^2} 
\Sigma^{\mathrm{fin}}_{\PAn \PZ} (\MA^2) \biggr]^{\, \mathrm{DCPR}}
\,=\, \tanb^{\overline{\mathrm{DR}}} \, ,
\end{equation}
where $\Sigma^{\mathrm{fin}}$ denotes the finite pieces of the
self-energy in the \DRbar\ scheme.  In this work, we always use the
renormalization condition (\ref{eq:wf_ren_A0}), also if we use
\DRbar~to renormalize $\tanb$. \end{itemize}

In analogy, there are the DCPR and the \DRbar~schemes for the
renormalization of the mass \MA~of the pseudoscalar Higgs boson. The
DCPR scheme uses the on-shell renormalization condition
\begin{equation}
\hat{\Sigma}_{\PAn} (\MA^2) \, = \, 0 \, ,
\end{equation}
while the \DRbar~scheme is again defined by setting
$\Sigma^{\mathrm{fin}}_{\PAn} (\MA^2)$  to zero in the mass
counterterm for the pseudoscalar Higgs boson. In this work, we use
the on-shell scheme for the renormalization of $\PAn$.
Supersymmetric models are usually defined in terms of \DRbar\
parameters~\cite{Aguilar-Saavedra:2005pw}. Hence, we have to
calculate the on-shell $\PAn$ mass from the corresponding
scale-dependent \DRbar\ parameter. For a given parameter set, we
determine $\MA^{\mathrm{os}}$ from the zero of the inverse $\PAn$
propagator%
\footnote{In contrast to \citere{Pierce:1996zz}, but in line with
  \citere{Aguilar-Saavedra:2005pw}, we assume that  all tadpole
  contributions to the mass of $\PAn$ are absorbed in the 
  definition of the \DRbar~mass $\MA^{\overline{\mathrm{DR}}}$.}
\begin{equation}
(\MA^{\mathrm{os}})^2 - 
(\MA^{\overline{\mathrm{DR}}})^2 + 
{\Sigma}^{\mathrm{fin}}_\PAn ((\MA^{\mathrm{os}})^2) = 0
\end{equation}
which corresponds to a given \DRbar\ mass for $\PAn$. Here,
${\Sigma}^{\mathrm{fin}}_\PAn (k^2)$ is calculated from an MSSM
parameter set using $\MA^{ \mathrm{os}}$ as input.  We start with
$\MA^{ \mathrm{os}} = \MA^{\overline{\mathrm{DR}}}$ for the
self-energy calculation and iterate until self-consistency is
reached. If $\tanb$ is renormalized in the DCPR scheme,
(\ref{eq:TB_conversion}) is used to also find $\tanb^{\,
\mathrm{DCPR}}$ self-consistently along with $\MA^{ \mathrm{os}}$.
The one-loop shift of the numerical value of $\MA$ is particularly
important because it enters already at tree level through the
determination of the mixing angle [see (\ref{eq:sa_tree_level})].

For completeness, we also state the remaining renormalization
conditions for the tadpoles,
\begin{equation}
T_\Phn + \delta t_\Phn \, = \, 0 \quad \mathrm{and} \quad
T_\PHn + \delta t_\PHn \, = \, 0 \, ,
\end{equation}
which ensure that $\vevu$ and $\vevd$ correctly minimize the
one-loop potential. As in the SM, the masses for the W and Z boson
are renormalized by on-shell conditions.

Including the corrections to the Higgs external legs, the partonic 
cross section to one-loop order is given by 
\begin{equation}
\begin{split}
\sigma^{(1)}_{\Phn} \, & = \, \sigma^{\mathrm{tree}}_{\Phn} \cdot
Z_\Phn \left[ \left( 1- Z_{\Phn \PHn} \frac{\cosa}{\sina} \right)^2 +
2 \, \Re \left( 1- Z_{\Phn \PHn} \frac{\cosa}{\sina} \right) \Delta_\Phn
\right] \, , \\
\sigma^{(1)}_{\PHn} \, & = \, \sigma^{\mathrm{tree}}_{\PHn} \cdot
Z_\PHn \left[ \left( 1- Z_{\PHn \Phn} \frac{\sina}{\cosa} \right)^2 +
2 \, \Re \left( 1 - Z_{\PHn \Phn} \frac{\sina}{\cosa} \right) \Delta_\PHn
\right] \, , \\
\sigma^{(1)}_{\PAn} \, & = \, \sigma^{\mathrm{tree}}_{\PAn} \cdot
Z_\PAn \left( 1 + 2 \, \Re \Delta_\PAn \right) \, , 
\end{split}
\end{equation} 
where $\Delta_{\PHgeneric}$ denotes the  relative one-loop vertex
corrections including the corresponding counterterms. Depending on
the  renormalization scheme, $\Delta_\PAn$ also includes the
contributions from $\PZ\PAn$-mixing and
$\mathswitchr{G^0}\PAn$-mixing.  The $Z$ factors are given
by~\cite{Dabelstein:1995js} 
\begin{equation} \label{eq:Zfacs1}
\begin{split}
Z_\Phn \, & = \left. \, \frac{1}{
1+ \hat{\Sigma}'_\Phn (k^2)- \left( 
\frac{\hat{\Sigma}^2_{\PHn\Phn}(k^2)}{
k^2-\MHntree^2+\hat{\Sigma}_\PHn (k^2)} \right)'} 
\right|_{k^2=M^2_\Phn} \, , \\
Z_\PHn \, & = \left.\, \frac{1}{
1+ \hat{\Sigma}'_\PHn (k^2)- \left( 
\frac{\hat{\Sigma}^2_{\PHn\Phn}(k^2)}{
k^2-\Mhntree^2+\hat{\Sigma}_\Phn (k^2)} \right)'} 
\right|_{k^2=M^2_\PHn} \, , \\
Z_\PAn \, & = \left.\, \frac{1}{1+ \hat{\Sigma}'_\PAn (k^2)} 
\right|_{k^2=M^2_\PAn} \,=\, 1 \, .
\end{split}
\end{equation}
The mixing of the CP-even Higgs bosons is determined by
\begin{equation} \label{eq:Zfacs2}
\begin{split}
Z_{\Phn \PHn} \, & = \, - \, \frac{\hat{\Sigma}_{\PHn\Phn}(\Mhn^2)}{
\Mhn^2 - \MHntree^2 + \hat{\Sigma}_{\PHn}(\Mhn^2)} \, , \\
Z_{\PHn \Phn} \, & = \, - \, \frac{\hat{\Sigma}_{\PHn\Phn}(\MHn^2)}{
\MHn^2 - \Mhntree^2 + \hat{\Sigma}_{\Phn}(\MHn^2)} \, ,
\end{split}
\end{equation}
where $m_{\PHgeneric}$ denotes the tree-level masses and
$M_{\PHgeneric}$ the  one-loop masses ($\PHgeneric = \Phn, \,
\PHn$), i.e.\ the zeros of the  inverse one-loop propagator matrix
determinant
\begin{equation} 
\label{eq:Higgs_masses}
\left( k^2 - \Mhntree^2 + \hat{\Sigma}_\Phn (k^2) \right) 
\left( k^2 - \MHntree^2 + \hat{\Sigma}_\PHn (k^2) \right) - 
\hat{\Sigma}^2_{\PHn\Phn} (k^2) \, = \, 0 \, .
\end{equation}
The renormalized Higgs self-energies in turn are given by 
\begin{equation} 
\begin{split} 
\hat{\Sigma}_\Phn (k^2) \, & = \, \Sigma_\Phn (k^2) + 
k^2 \, \left( \delta Z_{\PHu} \cosa^2 + \delta Z_{\PHd} \sina^2  \right)
- \delta \Mhntree^2 \, , \\
\hat{\Sigma}_\PHn (k^2) \, & = \, \Sigma_\PHn (k^2) + 
k^2 \, \left( \delta Z_{\PHu} \sina^2 + \delta Z_{\PHd} \cosa^2  \right)
- \delta \MHntree^2 \, , \\
\hat{\Sigma}_{\PHn \Phn} (k^2) \, & = \, \Sigma_{\PHn \Phn} (k^2) + 
k^2 \, \sina \cosa \left( \delta Z_{\PHu} - \delta Z_{\PHd} \right)
- \delta {\mathswitch {m^2_{\PHn \Phn}}} \, , \\
\end{split}
\end{equation} 
where the Higgs mass counterterms read%
\footnote{Up to some (known) sign errors in $\delta m_{\PHn
  \Phn}^2$, the Higgs mass counterterms can be also found 
  in~\citere{Dabelstein:1994hb}.}
\begin{eqnarray} 
\label{eq:higgsct}
\delta \Mhntree^2 \,& = & \, 
c^2_{\beta - \alpha} \, \Sigma_\PAn (\MA^2) + 
s^2_{\beta + \alpha} \Sigma_\PZ (\MZ^2) -
\frac{e \, s^2_{\beta - \alpha} \, c_{\beta - \alpha}}{2 \MZ \sw \cw} \, T_\PHn +
\frac{e \, s_{\beta - \alpha} \, (1 + c^2_{\beta - \alpha})}{2 \MZ \sw
  \cw} \, T_\Phn 
\nonumber \\[2ex]
&& \quad - \left( \MA^2 \, c^2_{\beta - \alpha} + \MZ^2 \, s^2_{\beta + \alpha} \right)
                                                            \Sigma'_\PAn (\MA^2) +
\MZ \, \frac{c_{2 \alpha} - c_{2 \beta}}{s_{2 \beta}} \, \Sigma_{\PAn \PZ}  (\MA^2) \, , \\[2ex]
\delta \MHntree^2 \,& = & \, 
s^2_{\beta - \alpha} \, \Sigma_\PAn (\MA^2) + 
c^2_{\beta + \alpha} \Sigma_\PZ (\MZ^2) +
\frac{e \, c_{\beta - \alpha} \, (1 + s^2_{\beta - \alpha})}{2 \MZ \sw \cw} \, T_\PHn -
\frac{e \, c^2_{\beta - \alpha} \, s_{\beta - \alpha}}{2 \MZ \sw \cw}
\, T_\Phn 
\nonumber \\[4ex]
&& \quad - \left( \MA^2 \, s^2_{\beta - \alpha} + \MZ^2 \, c^2_{\beta + \alpha} \right)
                                                            \Sigma'_\PAn (\MA^2) -
\MZ \, \frac{c_{2 \alpha} + c_{2 \beta}}{s_{2 \beta}} \, \Sigma_{\PAn
  \PZ} (\MA^2) \, , 
\\[2ex]
\delta m_{\PHn \Phn}^2 \,& = & \, 
- \frac{s_{2(\beta - \alpha)}}{2} \, \Sigma_\PAn (\MA^2) 
- \frac{s_{2(\beta + \alpha)}}{2} \, \Sigma_\PZ (\MZ^2) +
\frac{e \, s^3_{\beta - \alpha}}{2 \MZ \sw \cw} \, T_\PHn +
\frac{e \, c^3_{\beta - \alpha}}{2 \MZ \sw \cw} \, T_\Phn 
\nonumber \\[2ex]
&& \quad + \frac{1}{2} \, \left( \MA^2 \, s_{2(\beta - \alpha)} + 
                 \MZ^2 \, s_{2(\beta + \alpha)} \right)
                                                            \Sigma'_\PAn (\MA^2) 
+ \MZ \, \frac{s_{2 \alpha}}{s_{2 \beta}} \, \Sigma_{\PAn \PZ} (\MA^2) \, .
\end{eqnarray} 
If $\tanb$ is renormalized in the \DRbar\ scheme the finite parts of
$\Sigma_{\PAn \PZ}$ have to be set to zero in the above formulas. 
Note also that $\MA$ is the on-shell $\PAn$-mass in these formulas,
because we translate the $\MA$ input always to the on-shell scheme
before the actual loop calculation. As mentioned before, in all the
above equations we only consider the real parts of the
self-energies.  Neglecting the imaginary part does not spoil the
one-loop accuracy of our calculation, since the imaginary parts
formally only enter at higher orders.%
\footnote{While an inclusion of such imaginary parts in mass
  determinations of unstable particles is straightforward, a
  consistent inclusion of such width effects in the description of
  particle reactions requires an  inspection of full resonance
  processes including production and decay  of the unstable
  particles. This is beyond the aimed level of precision of this
  work.}

In complete analogy to SUSY-QCD in Section~\ref{se:nlo}, there are
$\tanb$-enhanced corrections in the electroweak sector which can be
numerically sizeable and which should be summed~\cite{Carena:1999py}
to all orders.  Higgsino--stop loops lead to a contribution to
$\Delta_\Pb$ of the form
\begin{equation}
\label{eq:ew_tanb_enhanced1}
\Delta_\Pb^{\tilde{\PH} \tilde{\Pt}} = 
\frac{\alpha \, \Mt^2}{8 \pi \sw^2 \, \sinb^2 \, \MW^2} 
A_\Pt \, \mu \, \tanb \, I(m_{\tilde{t}_1}, m_{\tilde{t}_2}, \mu) \, ,
\end{equation}
where $m^2_{\tilde{t}_i}$ denote the masses of the stop mass
eigenstates, and $A_\Pt$ is the usual trilinear soft breaking
parameter. From wino--higgsino--stop and wino--higgsino--sbottom
loops we find
\begin{equation}
\label{eq:ew_tanb_enhanced2}
\begin{split}
\Delta_\Pb^{\tilde{\PW}} = - \frac{\alpha}{8 \pi \sw^2}
M_2 \, \mu \, \, \tanb & \Bigl[ 
2\cos^2 \Theta_{\tilde{t}} \, \, I(m_{\tilde{t}_1},\mu,M_2) +
2\sin^2 \Theta_{\tilde{t}} \, \, I(m_{\tilde{t}_2},\mu,M_2)  \\
& + \cos^2 \Theta_{\tilde{\Pb}} \, \, I(m_{\tilde{\Pb}_1},\mu,M_2) +
    \sin^2 \Theta_{\tilde{\Pb}} \, \, I(m_{\tilde{\Pb}_2},\mu,M_2) 
\Bigr] \, ,
\end{split}
\end{equation}
where $M_2$ is the soft SUSY-breaking mass parameter for the wino,
$\Theta_{\tilde{t}}$ denotes the stop mixing angle, and the
auxiliary function $I(a,b,c)$ is given in (\ref{eq:I}). Finally,
bino loops contribute 
\begin{equation}
\label{eq:ew_tanb_enhanced3}
\begin{split}
\Delta_\Pb^{\tilde{\PB}} = - \frac{\alpha}{72 \pi \cw^2}
M_1 \, \mu \, \, \tanb & 
\Bigl[ 3(1+\sin^2 \Theta_{\tilde{\Pb}}) \, I(m_{\tilde{\Pb}_1},\mu,M_1) 
+3(1+\cos^2 \Theta_{\tilde{\Pb}}) \, I(m_{\tilde{\Pb}_2},\mu,M_1) \\
& + 2I(m_{\tilde{\Pb}_1},m_{\tilde{\Pb}_2},M_1)
\Bigr] \, ,
\end{split}
\end{equation} 
where $M_1$ is the soft SUSY-breaking mass parameter for the bino.
The full result for $\Delta_\Pb$ is then given by 
\begin{equation}\label{eq:deltab}
\Delta_\Pb = \Delta_\Pb^{\tilde{\Pg}} + \Delta_{\Pb}^{\mathrm{weak}} = 
\Delta_\Pb^{\tilde{\Pg}} + \Delta_\Pb^{\tilde{\PH} \tilde{\Pt}} +
\Delta_\Pb^{\tilde{\PW}} + \Delta_\Pb^{\tilde{\PB}},
\end{equation}
where $\Delta_\Pb^{\tilde{\Pg}}$ denotes the gluino contribution of
\refeq{eq:qcd_tanb_enhanced}.  We have recalculated these results
and find agreement with~\citere{Carena:1999py}.  In contrast
to~\citere{Carena:1999py} we also include the bino terms in the
summation. However, they are indeed numerically small.

To further improve the accuracy of the calculation, we also include
two-loop contributions to the self-energies as provided by  the
program package {\tt FeynHiggs}~\cite{FeynHiggs} (version~2.3.2). As
for the one-loop part of the self-energies we find perfect agreement
between our calculation and the results obtained with {\tt
FeynHiggs}.%
\footnote{Since {\tt FeynHiggs} does not directly support our 
  renormalization scheme ii) (the Higgs field renormalization is
  different),  we had to implement this scheme in {\tt FeynHiggs}
  ourselves. For the leading two-loop corrections included in 
  {\tt FeynHiggs}, neither $\tan \beta$ nor the Higgs fields are 
  renormalized at the two-loop level. Thus, it is consistent to 
  add the {\tt FeynHiggs} two-loop correction to the one-loop result
  in our renormalization schemes.} 

In addition to using a properly defined $\Pb$-mass, one often
absorbs parts of the radiative correction related to the Higgs
external leg using an effective mixing angle
$\alpha_\mathrm{eff}=\alpha+\delta\alpha$ at tree level. Here, we
follow {\tt FeynHiggs} and define $\delta\alpha$ to be the angle
which diagonalizes the loop-corrected Higgs mass matrix, i.e.
\begin{equation}
\label{eq:alpha_eff}
\left(
\begin{array}{cc}
\Mhntree^2 - \hat{\Sigma}_{\Phn}(\Mhntree^2) & - \hat{\Sigma}_{\Phn \PHn}((\Mhntree^2+\MHntree^2)/2)\\
- \hat{\Sigma}_{\Phn \PHn}((\Mhntree^2+\MHntree^2)/2) & \MHntree^2 - \hat{\Sigma}_{\PHn}(\MHntree^2) \\
\end{array}
\right)
\stackrel{\delta\alpha}{\longrightarrow}
\left(
\begin{array}{cc}
\Mhn^2 & 0 \\
0 & \MHn^2 \\
\end{array}
\right) ,
\end{equation}
where the renormalized Higgs self-energies are evaluated at the
given momenta.  We define an improved Born approximation
$\sigma_{\mathrm{IBA}}$ which includes the leading higher-order
corrections through the running $\Pb$-mass, the summation of
$\tanb$-enhanced terms and the effective mixing angle, 
\begin{equation}
\label{eq:iba}
\hat{\sigma}_{\rm {IBA}} \, = \, 
\frac{\sqrt{2}\pi G_\mu\overline{m}_{\Pb}(\mu_R)^2}{6\MHgeneric^2} \,
                    \, \delta(1-\tau) \, 
\left\{ \begin{array}{l} 
\frac{\displaystyle s^2_{\alpha_{\rm eff}}}{\displaystyle\cosb^2}
\left(\frac{\displaystyle 1 - \Delta_\Pb/(\tanb t_{\alpha_{\rm eff}})}{\displaystyle 1 + \Delta_\Pb}\right)^2 \, , \\[2ex]
\frac{\displaystyle c^2_{\alpha_{\rm eff}}}{\displaystyle\cosb^2}
\left(\frac{\displaystyle 1 + \Delta_\Pb \, t_{\alpha_{\rm eff}}/\tanb}{\displaystyle
    1 + \Delta_\Pb}\right)^2 \, , \\[2ex]
\tanb^2\left(\frac{\displaystyle 1 - \Delta_\Pb/\tanb^2}{\displaystyle
    1 + \Delta_\Pb}\right)^2 \end{array} 
\quad \mathrm{for} \, \, \, \, \begin{array}{l} \Phn \\[2.5ex] \PHn \\[2.5ex] \PAn \end{array} \, \, 
\mathrm{production},
           \right. 
\end{equation}
where $\hat{\sigma}_{\rm {IBA}}$ denotes the partonic cross section,
$\PHgeneric =(\Phn,\PHn,\PAn)$, $\tau = \MHgeneric^2/\hat{s}$, and
$\sqrt{\hat{s}}$ is the partonic CMS energy.  Here
$\overline{m}_{\Pb}(\mu_R)$ is the running \MSbar~bottom mass in
QCD, $\Delta_\Pb$ comprises the $\tanb$-enhanced terms of the
supersymmetric QCD and weak corrections as specified in
(\ref{eq:deltab}), and $\alpha_\mathrm{eff}$ is the effective mixing
angle defined in (\ref{eq:alpha_eff}).  We will compare our full
result to this approximation in Section~\ref{se:pheno}.

\section{Phenomenological analysis}
\label{se:pheno}

\subsection{SM input parameters}
\label{se:input}

For our numerical predictions, we essentially use the SM input
parameters~\cite{Yao:2006px}
\begin{equation}
\mbox{
\begin{tabular}{rclrclrcl}
$\alpha      $ & $ \! \! \! \! \! \! = \! \! \! \! \! \! $ & $ 1/137.03599911 $, & 
$\alpha(\MZ) $ & $ \! \! \! \! \! \! = \! \! \! \! \! \! $ & $ 1/128.952$, & 
$G_\mu       $ & $ \! \! \! \! \! \! = \! \! \! \! \! \! $ & $ 1.16637\times 10^{-5}~$GeV$^{-2}$, \\
$\MW         $ & $ \! \! \! \! \! \! = \! \! \! \! \! \! $ & $ 80.403$~GeV, & 
$\MZ         $ & $ \! \! \! \! \! \! = \! \! \! \! \! \! $ & $ 91.1876$~GeV, & & $\! \! \! \! \! \! $ &\\
$\Me         $ & $ \! \! \! \! \! \! = \! \! \! \! \! \! $ & $ 0.51099892$~MeV, & 
$\Mmy        $ & $ \! \! \! \! \! \! = \! \! \! \! \! \! $ & $ 105.658369$~MeV, & 
$\Mta        $ & $ \! \! \! \! \! \! = \! \! \! \! \! \! $ & $ 1.77699$~GeV, \\
$\Mu         $ & $ \! \! \! \! \! \! = \! \! \! \! \! \! $ & $ 66$~MeV, & 
$\Mc         $ & $ \! \! \! \! \! \! = \! \! \! \! \! \! $ & $ 1.2$~GeV,  & 
$\Mt         $ & $ \! \! \! \! \! \! = \! \! \! \! \! \! $ & $ 174.2$~GeV, \\
$\Md         $ & $ \! \! \! \! \! \! = \! \! \! \! \! \! $ & $ 66$~MeV, & 
$\Ms         $ & $ \! \! \! \! \! \! = \! \! \! \! \! \! $ & $ 0.15$~GeV, & 
$\overline{m}_{\Pb}(\overline{m}_{\Pb})    
             $ & $ \! \! \! \! \! \! = \! \! \! \! \! \! $ & $ 4.2$~GeV. \\
\end{tabular}
}
\end{equation}
Here, $\overline{m}_{\Pb}(\overline{m}_{\Pb})$ is the
QCD--\MSbar~mass for the bottom quark while the top mass $\Mt$ is
understood as an on-shell mass. For the QED renormalization of the
bottom mass we use the on-shell scheme, as mentioned above, with an
on-shell mass $\Mb = 4.53$~GeV calculated from
$\overline{m}_{\Pb}(\overline{m}_{\Pb})$ to one-loop order in QCD.
The masses of the light quarks are adjusted such as to reproduce the
hadronic contribution to the photonic vacuum polarization leading to
$\alpha(\MZ)$ of~\citere{Jegerlehner:2001wq}.  They are relevant
only for the evaluation of the charge renormalization constant
$\delta Z_e$ in the $\alpha(0)$-scheme. The CKM matrix has been set
to the unit matrix. For the calculation of the hadronic cross
sections we have adopted the MRST2004qed parton distribution
functions~\cite{Martin:2004dh} at NLO QCD and NLO QED, with the
corresponding $\alpha_{\rm s}(\MZ) =0.11899$. The top quark,
squarks, and gluinos are decoupled from the running of the strong
coupling $\alpha_{\mathrm{s}}(\mu_R)$. We choose the renormalization
and factorization scales as $\mu_R=\MH$ and $\mu_F=\MH/4$,
respectively. As mentioned above, with this specific choice the QCD
NNLO radiative corrections are at the percent
level~\cite{Harlander:2003ai}.

\subsection{The SM cross section}
\label{se:SM_results}
The NLO cross section predictions for associated $\Pb\bar{\Pb}\PH$
production at the LHC, including QCD, QED and weak corrections, are
shown in Table~\ref{tb:NLO_QCD_and_QED_cross_section} for the three
different input-pa\-ra\-me\-ter schemes introduced in
Section~\ref{se:ewcorr}.
\begin{table} 
\begin{center}
\begin{tabular}{|c|c|c|c|c|}
\hline
$\sigma$ [pb]        & LO      & QCD     & QCD+QED & full SM \\ \hline
$\alpha(0)$-scheme   & 0.02309 & 0.02868 & 0.02863 & 0.02901 {\small (25.6\%)} \\              
$G_{\mu}$-scheme     & 0.02390 & 0.02968 & 0.02963 & 0.02924 {\small (22.3\%)} \\
$\alpha(\MZ)$-scheme & 0.02453 & 0.03047 & 0.03042 & 0.02932 {\small (19.5\%)} \\ \hline
\end{tabular}
\end{center}
\mycaption{\label{tb:NLO_QCD_and_QED_cross_section} The LO and NLO
 SM cross section $\Pp\Pp\to (\Pb\bar{\Pb})\,\PH\!+\!X$ for a  Higgs
 boson with $\MH~=~300$~GeV at the LHC ($\sqrt{s}=14$~TeV).  Results
 are presented for the three different input-parameter schemes
 defined in Section~\ref{se:ewcorr}. The MRST2004qed parton 
 distribution function and NNLO-QCD running for the $\Pb$-mass have
 been  adopted for the NLO as well as the LO cross sections, and the
 renormalization and factorization scales have been set to $\mu_R =
 \MH$ and $\mu_F = \MH/4$, respectively. ``QCD''  denotes the NLO
 QCD corrections only, ``QCD+QED'' also includes photon exchange and
 emission as well as the initial state containing a photon. The
 ``full SM'' prediction includes all ${\cal O}(\alpha_s)$ and ${\cal
 O}(\alpha)$ corrections; the corresponding relative correction is
 indicated in parentheses.}
\end{table}
The QED corrections are very small after potentially large
contributions from collinear photon emission have been removed by
mass factorization as described in Section~\ref{se:ewcorr}. As
expected, the inclusion of the electroweak corrections reduces the
scheme dependence. In the following we will adopt the
$G_\mu$-scheme, where the value of the electromagnetic coupling is
derived from muon decay according to $\alpha_{G_\mu}$ and where the
influence of the light-quark masses is negligible. Note that the
relative ${\cal O}(\alpha)$ QED corrections are evaluated with
$\alpha(0)$, irrespective of the chosen input-parameter scheme,
because the relevant scale for the bremsstrahlung process is set by
the vanishing virtuality of the emitted real photon.

\subsection{MSSM input}
\label{se:MSSM_input}

For our numerical analysis, we will focus on the benchmark scenario
SPS~4~\cite{Allanach:2002nj} which is cha\-rac\-terized by a large
value of $\tanb = 50$ and a correspondingly large associated
production cross section $\Pb\bar{\Pb}\to \PHn,\PAn$ at the LHC. 
Results for the alternative benchmark scenario SPS~1b with $\tanb
=30$ will be presented in Appendix~\ref{app:SPS1b}. Both the SPS~4
and SPS~1b input parameters are specified in
Appendix~\ref{app:SPS}.  Note that searches at the Fermilab Tevatron
are not sensitive to scenarios with $\PHn,\PAn$-masses as large as
the masses in SPS~4 or SPS~1b.

From the SPS \DRbar\ input parameters we calculate
$\MA^{\mathrm{os}}$ and, if we work in the DCPR scheme, also
$\tanb^{\mathrm{DCPR}}$, as specified in
Section~\ref{se:ewmssmcorr}. The corresponding renormalization scale
$\mu_R$(\DRbar)  used in the electroweak part of the calculation is
specified by the SPS scenario (see Appendix~\ref{app:SPS}). Note,
that it differs in general from the renormalization scale in the QCD
part of the calculation which is set to the mass of the produced
Higgs boson.  The MSSM tree-level relations are used to determine
the sfermion and gaugino masses and mixing angles that enter the
one-loop corrections.  The Higgs masses and the Higgs sector mixing
angle $\alpha$ [see (\ref{eq:sa_tree_level})] which enter the
calculation of the loop diagrams are also obtained according to the
tree-level relations.

While the bottom mass that enters the Yukawa coupling at LO is fixed
by the requirement to account for dominant NLO corrections, 
different definitions for the bottom mass in the relative NLO
correction change the result only beyond NLO.  As argued in
Section~\ref{se:ewcorr}, we shall use the running bottom mass as
defined after summation of $\tanb$-enhanced terms as input%
\footnote{While the electroweak corrections are calculated  using
  dimensional reduction, we do not convert the running  \MSbar\
  bottom mass to the corresponding \DRbar\ mass to define the input
  value for the relative one-loop correction. The difference is  of
  higher order.}.
The running mass is needed as an input for the determination of the
MSSM parameters, but it also depends on these parameters through the
QCD renormalization scale $\mu_R = \MHgeneric$ and through the
$\tanb$-enhanced corrections. The $\Pb$-mass is thus calculated
using an iterative procedure,  starting from some initial guess for
its value, and using the resulting $\Pb$-mass as input for a refined
determination of the MSSM parameters, until self-consistency is
reached. Note, however, that this running mass depends on the
process under consideration through the choice of scale and through
the $\tanb$-enhanced corrections. In order to avoid the
proliferation of input masses, we adopt the running $\Pb$-mass
associated with $\PHn$ production in the relative corrections to all
processes, e.g.\  $\Mb=2.24$~GeV for SPS~4 in the \DRbar\ scheme.

We use the two-loop-improved Higgs masses for the kinematics, e.g.\
in the tree-level cross section or when they appear as external
momenta in the on-shell vertex-correction. Using the two-loop
self-energies from {\tt FeynHiggs}~\cite{FeynHiggs},  these two-loop
improved on-shell Higgs-boson masses in the SPS~4~scenario are given
by $\Mhn = 115.66$~GeV, $\MHn = 397.72$~GeV, and $ \MA = 397.67$~GeV
for the renormalization scheme~ii) introduced in
Section~\ref{se:ewmssmcorr}, i.e.\ what we call  \DRbar~scheme.  (As
described in Section~\ref{se:ewcorr} we only consider the real part
of the self-energies when we determine the Higgs masses, in contrast
to the default setting in {\tt FeynHiggs}. Including the imaginary
parts has no visible effect on $\Mhn$ and shifts $\MHn$ by only
approximately 200~MeV.)

\subsection{The MSSM cross sections}
\label{se:MSSM_results}

Within the MSSM, let us first focus on the radiative corrections and
total cross sections in the SPS~4 benchmark scenario using the
\DRbar\ scheme for the renormalization of $\tanb$.

In Table~\ref{Tab:results_SPS4} we present the relative radiative
corrections $\delta$ defined with respect to the improved Born
approximation $\sigma_{\rm {IBA}}$ defined in Eq.~\refeq{eq:iba}. 
\begin{table}
\begin{center}
\begin{tabular}{|c|r|r|r|}
\hline
& \multicolumn{1}{c|}{$\Phn$} & \multicolumn{1}{c|}{$\PHn$} & 
\multicolumn{1}{c|}{$\PAn$} \\ \hline
$\sigma_{\mathrm{IBA}}$[pb] &          0.65 &    15.39 &    15.40 \\ \hline
$\delta_{\mathrm{QCD}}[\%]$ &         36.25 &    21.48 &    21.48 \\ 
$\delta_{\mathrm{QED}}[\%]$ &         -0.13 &    -0.23 &    -0.23 \\ 
$\delta_{\mathrm{MSSM-QCD}}[\%]$ &    -0.03 &     0.08 &     0.07 \\ 
$\delta_{\mathrm{MSSM-weak}}[\%]$ &   -1.22 &    -1.57 &    -1.60 \\ 
\hline
\end{tabular}
\end{center}
\mycaption{\label{Tab:results_SPS4}  LO cross section in the
  improved Born approximation $\sigma_{\mathrm{IBA}}$ as defined in
  (\ref{eq:iba}), as well as  corrections $\delta$ relative to
  $\sigma_{\mathrm{IBA}}$, for  $\Pp\Pp\to
  (\Pb\bar{\Pb})\,\Phn/\PHn/\PAn\!+\!X$ at the LHC
  ($\sqrt{s}=14$~TeV) in the SPS~4 scenario.  The MRST2004qed PDFs
  and NNLO-QCD running for $\Mb$ have been adopted,  the
  renormalization and factorization scales have been set to  $\mu_R
  = \MHgeneric$ and $\mu_F = \MHgeneric/4$, respectively, and
  $\tanb$ has been renormalized in the \DRbar~scheme. ``QCD''
  denotes the NLO QCD corrections only, ``QED'' denotes all photonic
  corrections only, and ``MSSM-QCD'' and ``MSSM-weak'' comprise only
  the QCD and weak effects in the MSSM, respectively, that remain
  after absorbing the  large-$\tanb$ effects in the LO cross
  section.}
\end{table}
The full cross-section prediction including summations and the
remaining non-universal ${\cal O}(\alpha_s)$ and ${\cal O}(\alpha)$
corrections is thus given by $\sigma = \sigma_{\rm {IBA}}\times
\left( 1+ \delta_{\rm QCD}+\delta_{\rm QED}\right.$ $\left.
+\delta_{\rm MSSM-QCD} + \delta_{\rm MSSM-weak}\right)$. Note that
we have removed corrections from our full calculation that are taken
into account through the use of $\alpha_\mathrm{eff}$ in
$\sigma_{\rm {IBA}}$ to avoid double counting.  As can be seen from
Table~\ref{Tab:results_SPS4}, the bulk of the MSSM-QCD and -weak
corrections can indeed be absorbed into the above definition of
$\sigma_{\rm {IBA}}$. The remaining non-universal corrections
$\delta_{\rm MSSM} = \delta_{\mathrm{MSSM-QCD}} +
\delta_{\mathrm{MSSM-weak}}$ in the complete MSSM calculation turn
out to be quite small, below approximately 2\%.  We note, however,
that the size of the corrections $\delta_{\rm MSSM}$ depends quite
sensitively on the numerical value of the input $\Pb$-mass. We shall
discuss this point in more detail below.

In Table~\ref{Tab:results_SPS4_cross_sections}, we show the
cumulating effect of the various higher-order corrections on the
effective  $\Pb$-mass (as defined after summation of
$\tanb$-enhanced terms,  $\Mb \! \to \!
\Mb(1-\Delta_\Pb/(\tanb\tana))/(1~+~\Delta_\Pb)$ for
$\Phn$~production, etc.) and the resulting cross sections. 
\begin{table}
\begin{center}
\begin{tabular}{|l|c|c|c|c|c|c|}
\hline
& \multicolumn{2}{|c|}{$\Phn$} & 
\multicolumn{2}{|c|}{$\PHn$} &
\multicolumn{2}{|c|}{$\PAn$} \\ \cline{2-7}
&  $\Mb$[GeV]  & $\sigma$[pb] & 
 $\Mb$[GeV]  & $\sigma$[pb] & 
 $\Mb$[GeV]  & $\sigma$[pb] \\ \hline
QCD &                                  2.80 &     0.97 &     2.55 &    24.12 &     2.55 &    24.13 \\ 
+QED &                                 2.80 &     0.97 &     2.55 &    24.07 &     2.55 &    24.09 \\ 
$+\Delta_{\Pb}^{\tilde{g}}$ &          2.72 &     0.92 &     1.95 &    14.14 &     1.95 &    14.15 \\ 
$+\Delta_{\Pb}^{\mathrm{weak}}$ &      2.75 &     0.94 &     2.24 &    18.66 &     2.24 &    18.67 \\ 
$+ \sin (\alpha_{eff})$ &              2.75 &     0.88 &     2.24 &    18.66 &     2.24 &    18.67 \\ \hline
full calculation &                     2.75 &     0.87 &     2.24 &    18.43 &     2.24 &    18.44 \\ 
\hline
\end{tabular}
\end{center}
\mycaption{\label{Tab:results_SPS4_cross_sections} The NLO MSSM
  cross section $\Pp\Pp\to (\Pb\bar{\Pb})\,\Phn/\PHn/\PAn\!+\!X$ at
  the LHC ($\sqrt{s}=14$~TeV) in the SPS~4 scenario including the
  cumulative corrections due to the different classes of corrections
  (PDFs, scale setting, and $\tanb$ renormalization as in 
  \refta{Tab:results_SPS4}). We also quote the effective bottom
  mass  entering the respective effective Yukawa coupling, to
  quantify the impact of the summation.  ``QCD'' denotes the NLO QCD
  corrections only, ``QED'' denotes the addition of all photonic
  corrections, $\Delta_{\Pb}^{\tilde{g}}$  and
  $\Delta_{\Pb}^{\mathrm{weak}}$ refer to the effect of summing
  $\tanb$ enhanced  terms in SUSY-QCD and the weak MSSM,
  respectively, $\sin (\alpha_{\rm eff})$  denotes the additional
  improvement of the Born cross section by using the  effective
  mixing angle $\alpha_{\rm eff}$. Finally, we give the result for
  the full calculation in the MSSM where all ${\cal O}(\alpha_s)$
  and  ${\cal O}(\alpha)$ corrections are included.}
\end{table}
As already observed within the SM, the QED corrections are generally
very small after mass factorization. The summation of the
$\tanb$-enhanced MSSM-QCD and MSSM-weak corrections, encoded in
$\Delta_{\Pb}^{\tilde{g}}$ and $\Delta_{\Pb}^{\mathrm{weak}}$,
respectively, has a significant effect on the cross sections for
$\PHn$ and $\PAn$ production. The light Higgs boson $\Phn$, on the
other hand, is SM-like in the SPS~4 scenario and the summation of
terms $\propto \tanb$ has thus no sizeable impact on the cross
section. Employing a loop-improved effective mixing angle
$\alpha_{\rm eff}$ is numerically relevant only for $\Phn$
production because $\sina \sim -1/\tanb$ is small and even a small
shift $\alpha \to \alpha_{\rm eff}$ has a sizeable effect on
$\sina$.   The cross sections in the last-but-one row of 
Table~\ref{Tab:results_SPS4_cross_sections} correspond to the
improved Born approximation dressed with QCD and QED corrections.
The full MSSM cross section including all summations and the
remaining non-universal ${\cal O}(\alpha_s)$ and ${\cal O}(\alpha)$
corrections is displayed in the last row of the table. As discussed
previously, the non-universal supersymmetric corrections turn out to
be small at a level of a few percent. Note that the $\Pb$-mass
values in the last row of \refta{Tab:results_SPS4_cross_sections}
have been used to calculate the cross sections tabulated in
\refta{Tab:results_SPS4}. (While $m_\Pb=2.75$~GeV and
$m_\Pb=2.24$~GeV have been used to calculate $\sigma_{\mathrm{IBA}}$
for $\Phn$ and $\PHn/\PAn$ production, respectively,
$m_\Pb=2.24$~GeV has been used as input for all relative
corrections, as discussed before.)

\begin{figure}
\begin{center}
\myplot{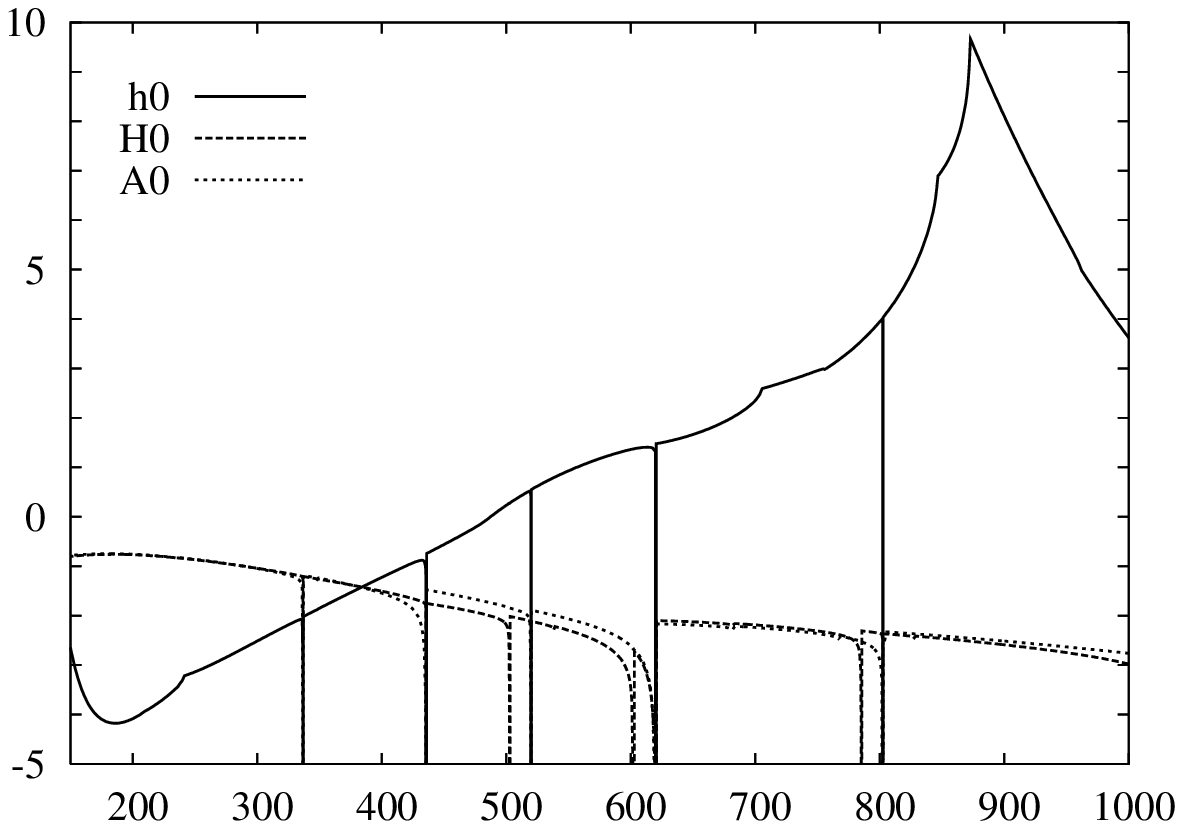}{146}{-9}{$\MA$[GeV]}{-30}{100}{
$\delta_{\mathrm{MSSM}}$[\%]}{0}{10cm}
\end{center}
\mycaption{\label{Fig:MA_plot} Full MSSM corrections $\delta_{\rm
    MSSM} = \delta_{\mathrm{MSSM-QCD}} + \delta_{\mathrm{MSSM-weak}}$
  defined relative to $\sigma_{\mathrm{IBA}}$ as a function of the
  $\MA$ pole mass in the \DRbar\ scheme for $\tanb$.  All other MSSM
  parameters are fixed to their SPS~4 values.}
\end{figure}

In Fig.~\ref{Fig:MA_plot} we show the impact of the complete
supersymmetric ${\cal O}(\alpha_s)$ and ${\cal O}(\alpha)$
corrections, $\delta_{\rm MSSM} = \delta_{\mathrm{MSSM-QCD}} +
\delta_{\mathrm{MSSM-weak}}$, defined relative to the improved Born
approximation $\sigma_{\rm IBA}$ (\ref{eq:iba}), for different
values of the on-shell mass $\MA$.  
\begin{figure}
\begin{center}
\myplot{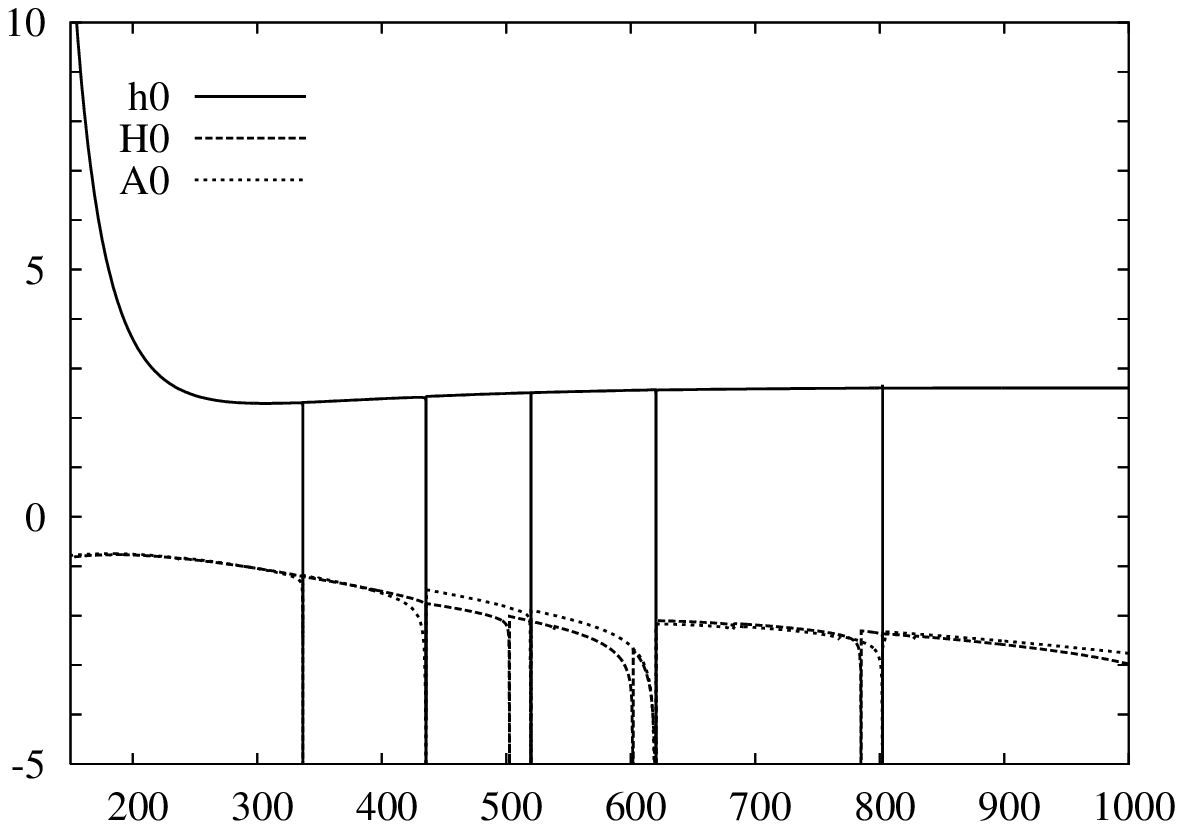}{146}{-9}{$\MA$[GeV]}{-30}{100}{
$\delta_{\mathrm{MSSM}}$[\%]}{0}{10cm}
\end{center}
\mycaption{\label{Fig:MA_plot_p2_at_0} Full MSSM corrections $\delta_{\rm
    MSSM} = \delta_{\mathrm{MSSM-QCD}} + \delta_{\mathrm{MSSM-weak}}$
    as in Fig.~\ref{Fig:MA_plot}. However, here 
    $\alpha_{\rm eff}$~(\ref{eq:alpha_eff}) in 
    $\sigma_{\mathrm{IBA}}$ is calculated from self-energies at $p^2=0$.}
\end{figure}
All other MSSM parameters are kept fixed to the SPS~4 values. The size
of the non-universal corrections does not exceed $3\%$ for $\PHn/\PAn$
production except for special model parameters, where the Higgs masses
are close to the production threshold for pairs of sparticles.  The
peaks in the corrections correspond to neutralino, chargino, or
sfermion thresholds. Thresholds which are too narrow, however, are not
displayed in Fig.~\ref{Fig:MA_plot}. These unphysical singularities
can be removed by taking into account the finite widths of the
unstable sparticles (see e.g.\ \citere{Bhattacharya:1991gr}).  Note
that the peaks for $\Phn$ production are induced by the finite parts
of the counterterms in~(\ref{eq:wf_ren_A0}). This proliferation of
unphysical singularities can be avoided by choosing the \DRbar\ scheme
for the $\PAn$ wave function renormalization, as is default in {\tt
  FeynHiggs}.

For small $\MA$ when the masses of the neutral Higgs bosons are
almost degenerate, the effects from the loop-induced Higgs mixing
become extremely large. To go beyond the effective mixing angle
approximation in this region, one would have to include corrections
to the off-shell $\Pb\Pbbar\Phn/\Pb\Pbbar\PHn$ vertex for
$\PHn/\Phn$ production, respectively.  We will not address this
issue, which is part of a two-loop calculation, in this work.
Therefore, we truncate Fig.~\ref{Fig:MA_plot} at $\MA = 150$~GeV.

For very large $\MA$ the relative corrections to $\Phn$
production increase to up to 10\%. However, in this parameter
region, the size of $\delta_{\mathrm{MSSM}}$ for $\Phn$ depends
very sensitively on the definition of the effective mixing angle
$\alpha_{\rm eff}$ employed in $\sigma_{\rm IBA}$. As defined 
in~(\ref{eq:alpha_eff}), $\alpha_{\rm eff}$ inherits a distinct 
dependence on $\MA$ because some of the self-energies are 
evaluated at momentum scales of the order of $\MHn \sim \MA$. 
This dependence is not present in the complete result and, thus, 
has to be compensated by $\delta_{\mathrm{MSSM}}$. Evaluating the
self-energies that enter $\alpha_{\rm eff}$~(\ref{eq:alpha_eff})
at $p^2=0$, the peak structure for large $\MA$ in
Fig.~\ref{Fig:MA_plot} is absent, and the size of the correction
is~2--3\%  for 300~GeV~$\lsim~\MA~\lsim$~1000~GeV,  as shown in
Fig.~\ref{Fig:MA_plot_p2_at_0}. However, this  approximation
breaks down at small values of $\MA$ where the corrections due to
Higgs mixing become large. Note that in any case
$\Phn$ is SM-like at large $\MA$ so that $\Phn$ production is
most likely of no phenomenological relevance.

The two-loop improvement of the self-energies of the CP-even Higgs
bosons, as contained in the $Z$~factors of \refeq{eq:Zfacs1} and
\refeq{eq:Zfacs2}, has a negligible effect on $\delta_{\rm MSSM}$
for $\PHn$ and $\PAn$ production. Only for $\Phn$ production and
$\MA \lsim\ 350$~GeV the difference between a one-loop and two-loop
treatment is larger than $1\%$ and can reach the 5\% level for  $\MA
\approx 200$~GeV.  However, the two-loop improvement plays an
important role for the precise determination of the Higgs-boson pole
masses entering the kinematics of the
process~\cite{Degrassi:2002fi}. 

\begin{figure}[fp]
\begin{center}
\myplot{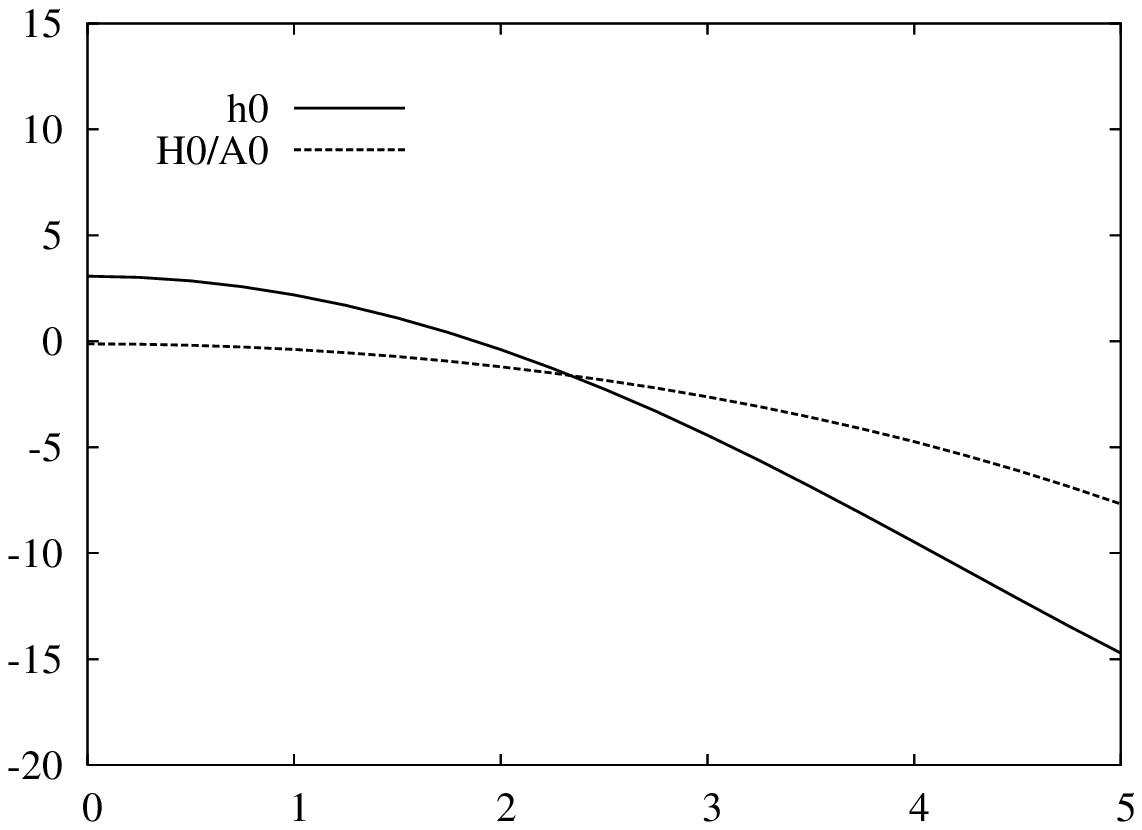}{135}{-9}{$\Mb$[GeV]}{-30}{100}{
$\delta_{\mathrm{MSSM}}$[\%]}{0}{10cm}\\[10mm]
\myplot{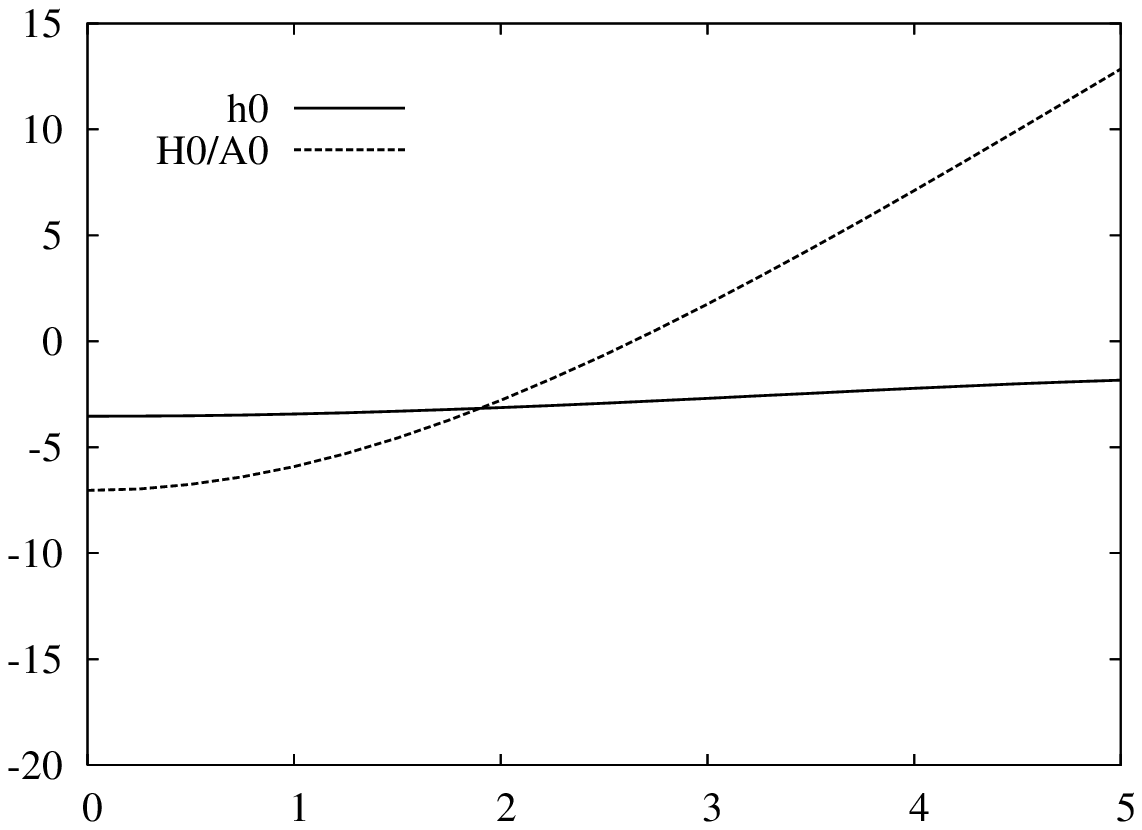}{135}{-9}{$\Mb$[GeV]}{-30}{100}{
$\delta_{\mathrm{MSSM}}$[\%]}{0}{10cm}
\begin{picture}(0,0)
\put(-240,267){\large $\tanb^{\mbox{\footnotesize \DRbar}}$}
\put(-240,40){\large $\tanb^{\mathrm{DCPR}}$}
\end{picture}
\end{center}
\mycaption{\label{Fig:MB_plot} Full MSSM corrections $\delta_{\rm
  MSSM} = \delta_{\mathrm{MSSM-QCD}} + \delta_{\mathrm{MSSM-weak}}$
  defined relative to $\sigma_{\mathrm{IBA}}$ as a function of the
  $\Mb$ input in the \DRbar\ scheme (upper panel) and DCPR scheme
  (lower panel) for $\tanb$.  The corrections for $\PHn$ and $\PAn$
  lie on top of each other.}
\end{figure}

It is important to emphasize that the size of the non-universal MSSM
one-loop corrections $\delta_{\mathrm{MSSM}}$ at large $\tanb$ is
quite sensitive to the choice of the bottom-mass input value. This
is caused by terms that grow as $\Mb^2\, \tanb^2$ but are not
included in the summation of $\tanb$-enhanced terms.  For the SPS~4
scenario, the $\Pb$-mass input sensitivity is shown in
Fig.~\ref{Fig:MB_plot}. 
In the \DRbar\ scheme for $\tanb$ (upper panel of
Fig.~\ref{Fig:MB_plot}) the absolute size of the non-universal
corrections varies between approximately zero and $-6\%$  for
$\PHn/\PAn$ production, depending on whether a massless
approximation, the running mass or the pole mass is chosen as
$\Pb$-mass input. The sensitivity is even larger for the light Higgs
$\Phn$. However, for $\Phn$ production $\delta_{\mathrm{MSSM}}$
depends very sensitively on the definition of $\alpha_{\rm eff}$, as
mentioned before. Although we assume that the running mass including
the summation of $\tanb$-enhanced terms is a sensible choice, the
sensitivity of the NLO correction to the $\Pb$-mass input
constitutes a theoretical uncertainty which can not be resolved at
the NLO level.

Comparing the \DRbar\ and the DCPR schemes for the renormalization
of $\tanb$, the size of the corrections is accidentally very similar
if the running $\Pb$-mass is chosen as an input. The same
observation holds for the Higgs masses, which differ by less than
500~MeV. However, because of the large $\Mb$ dependence the
corrections in the different renormalization schemes can differ
significantly in general. This can be seen by comparing the upper
and lower panels of Fig.~\ref{Fig:MB_plot}. For the Higgs masses we
also find larger differences up to 4~GeV for $\Mb = 4.2$~GeV. Of
course, the large scheme difference of the relative correction is
partially compensated by the corresponding change in the DCPR input
value for $\tanb$ to be calculated from the \DRbar\ value
$\tanb=50$. In the massless-$\Pb$ approximation, we find  $\tanb =
51.78$, while $\tanb = 47.00$ for $\Mb=4.2$~GeV. The resulting
scheme dependence of the total cross-section prediction for $\PHn$
and $\PAn$ production is thus moderate (below 1\%) even for large
$\Mb$. For $\Phn$ and large $\Mb$,  the residual scheme dependence
can reach up to 3\%. To compare the  cross sections, we have used
the on-shell mass for $\PAn$ as  computed in the \DRbar~scheme as
input in both schemes.

We have verified that the size of $\delta_{\rm MSSM}$ for $\PHn$ or
$\PAn$ production does not change significantly when the input
values of the soft breaking parameters for the sfermions or the soft
breaking parameters for the gauginos are varied around their SPS~4
values by up to a factor of 2, unless some sparticles become
unreasonably light.

Results for the alternative benchmark scenario SPS~1b with $\tanb
=30$ are presented in Appendix~\ref{app:SPS1b}. The conclusions
reached for the SPS~4 scenario essentially hold for SPS~1b as well.

\section{Conclusions}
\label{se:conclusion}

We have performed a complete calculation of the ${\cal O}(\alpha_s)$
and ${\cal O(\alpha)}$ corrections to associated bottom--Higgs
production through $\Pb\bar \Pb$ fusion in the MSSM.  This
next-to-leading order prediction is improved by including two-loop
corrections, as provided by {\tt FeynHiggs}, and by the known
summation of $\tanb$-enhanced corrections. We have presented
details of the calculation and discussed numerical results for MSSM
Higgs-boson production at the LHC in two supersymmetric scenarios.
The leading supersymmetric higher-order corrections, in particular
the $\tanb$-enhanced contributions, can be taken into account by
an appropriate definition of the couplings and the running
$\Pb$-mass in an improved Born approximation. The remaining
non-universal corrections are small, typically of the order of a few
percent.  The quality of an improved Born approximation, however,
can only be judged with a full ${\cal O}(\alpha_s)$ and ${\cal
O(\alpha)}$ calculation. Although we assume that the running mass
including the summation of $\tanb$-enhanced terms is a sensible
choice for the input $\Pb$-mass, alternative choices can lead to
non-universal corrections as large as 10\%. 

With the results presented, the impact of the complete MSSM
corrections to neutral Higgs boson decays into bottom quarks might
also be updated.

Our results show that the difference between a properly defined
improved Born approximation and the complete NLO calculation, which
is improved by leading NNLO effects, is smaller than other
theoretical uncertainties resulting from residual scale dependences,
errors on the b-quark mass, and parton distribution functions.  

\section*{Acknowledgments}
We are grateful to Sven Heinemeyer, Wolfgang Hollik, and Georg
Weiglein for discussions and comments on the manuscript.

\begin{appendix}

\section{SPS benchmark scenarios}
\label{app:SPS}

For the SPS benchmark~\cite{Allanach:2002nj} scenarios discussed in
this work we use the following input for $\tanb$, the mass of the
CP-odd Higgs boson $\MA$, the supersymmetric Higgs mass parameter
$\mu$, the electroweak gaugino mass parameters $M_{1,2}$, the gluino
mass $m_{\tilde{g}}$, the trilinear couplings $A_{\tau,\Pt,\Pb}$,
the scale, at which the \DRbar\ input values are defined $\mu_R
($\DRbar$)$, the soft SUSY-breaking parameters in the diagonal
entries of the squark and slepton mass matrices of the first and
second generation $M_{fi}$ (where $i=L,R$ refers to the left- and
right-handed sfermions, $f=q,l$ to quarks and leptons, and $f=u,d,e$
to up and down quarks and electrons, respectively), and the
analogous soft SUSY-breaking parameters for the third generation
$M^{3G}_{fi}$:
\begin{center}
\begin{tabular}{|r|r|r|}
\hline
\mbox{} \hspace{1cm} \mbox{} & SPS~4 & SPS~1b \\ \hline
$\tanb$            & $  50.0$ & $  30.0$ \\
$\MA$[GeV]              & $ 404.4$ & $ 525.5$ \\
$\mu$[GeV]              & $ 377.0$ & $ 495.6$ \\
$M_1$[GeV]              & $ 120.8$ & $ 162.8$ \\
$M_2$[GeV]              & $ 233.2$ & $ 310.9$ \\
$m_{\tilde{g}}$[GeV]    & $ 721.0$ & $ 916.1$ \\
$A_\tau$[GeV]           & $-102.3$ & $-195.8$ \\
$A_\Pt$[GeV]              & $-552.2$ & $-729.3$ \\
$A_\Pb$[GeV]              & $-729.5$ & $-987.4$ \\
$\mu_R ($\DRbar$)$[GeV] & $ 571.3$ & $ 706.9$ \\ \hline
\end{tabular}
\hspace{1cm}
\begin{tabular}{|r|r|r|}
\hline
\mbox{} \hspace{1.2cm} \mbox{} & SPS~4 & SPS~1b \\ \hline
$M_{qL}$[GeV]           &  $732.2$ &  $836.2$ \\
$M_{dR}$[GeV]           &  $713.9$ &  $803.9$ \\
$M_{uR}$[GeV]           &  $716.0$ &  $807.5$ \\
$M_{lL}$[GeV]           &  $445.9$ &  $334.0$ \\
$M_{eR}$[GeV]           &  $414.2$ &  $248.3$ \\
$M^{3G}_{qL}$[GeV]      &  $640.1$ &  $762.5$ \\
$M^{3G}_{dR}$[GeV]      &  $673.4$ &  $780.3$ \\
$M^{3G}_{uR}$[GeV]      &  $556.8$ &  $670.7$ \\
$M^{3G}_{lL}$[GeV]      &  $394.7$ &  $323.8$ \\
$M^{3G}_{eR}$[GeV]      &  $289.5$ &  $218.6$ \\ \hline
\end{tabular}
\end{center}

\section{Results for SPS 1b}
\label{app:SPS1b}

\begin{figure}
\begin{center}
\myplot{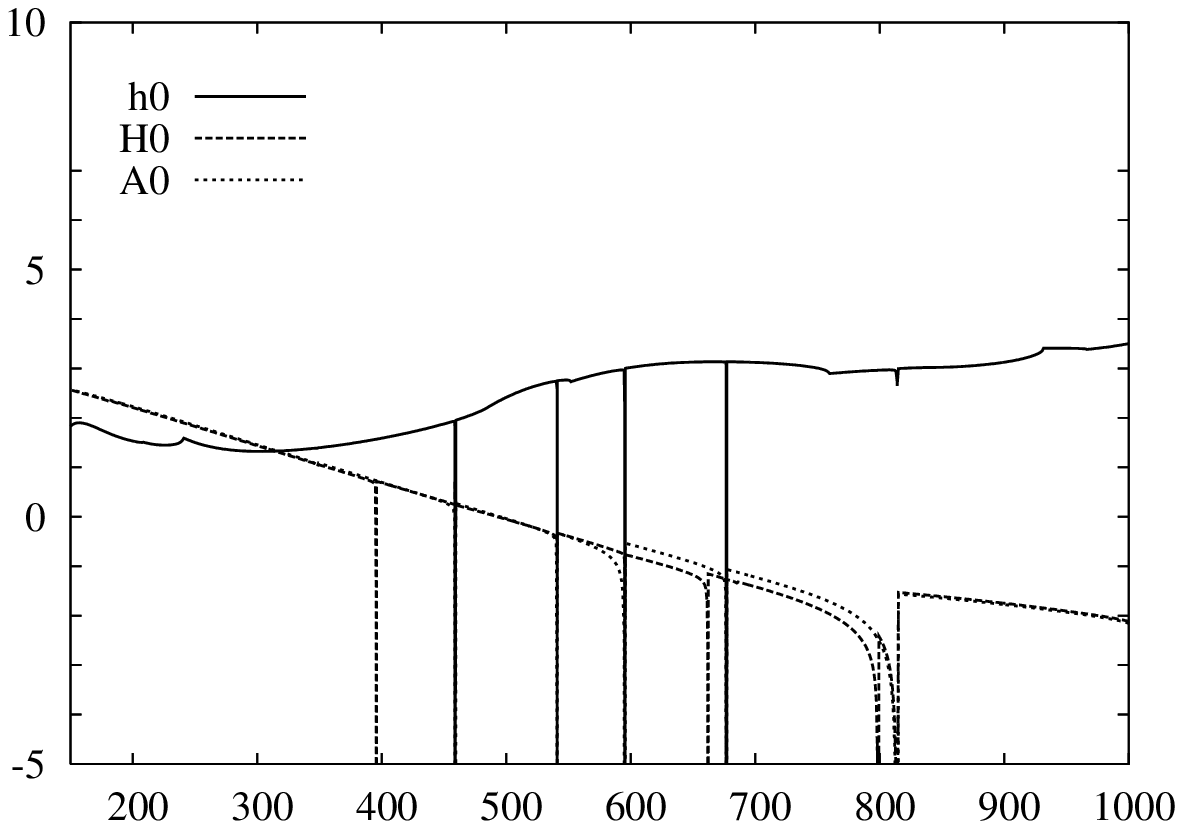}{146}{-9}{$\MA$[GeV]}{-30}{100}{
$\delta_{\mathrm{MSSM}}$[\%]}{0}{10cm}
\end{center}
\mycaption{\label{Fig:MA_plot_SPS1b} Full MSSM corrections $\delta_{\rm
    MSSM} = \delta_{\mathrm{MSSM-QCD}} + \delta_{\mathrm{MSSM-weak}}$
  defined relative to $\sigma_{\mathrm{IBA}}$ as a function of the
  $\MA$ pole mass in the \DRbar\ scheme for $\tanb$.  All other MSSM
  parameters are fixed to their SPS~1b values.}
\end{figure}

In the SPS~1b scenario the two-loop Higgs masses are given by  $\Mhn
= 117.67$~GeV, $\MHn = 525.69$~GeV, and $ \MA = 525.66$~GeV in the
\DRbar~scheme for $\tanb$. Here, the masses in the DCPR scheme also
differ by not more than 500~MeV and the discrepancy does not
increase as drastically as in SPS~4 when the input value for the
$\Pb$-mass is changed with respect to its default value $\Mb =
2.30$~GeV.   As can be seen from Table~\ref{Tab:results_SPS1b}, the
non-universal corrections in the MSSM for $\PAn/\PHn$-production are
even smaller than for SPS~4.
\begin{table}[b]
\begin{center}
\begin{tabular}{|c|r|r|r|}
\hline
& \multicolumn{1}{c|}{$\Phn$} & \multicolumn{1}{c|}{$\PHn$} & 
\multicolumn{1}{c|}{$\PAn$} \\ \hline
$\sigma_{\mathrm{IBA}}$[pb] &         0.59 &     1.88 &     1.88 \\ \hline
$\delta_{\mathrm{QCD}}[\%]$ &        35.98 &    19.02 &    19.02 \\ 
$\delta_{\mathrm{QED}}[\%]$ &        -0.13 &    -0.26 &    -0.26 \\ 
$\delta_{\mathrm{MSSM-QCD}}[\%]$ &   -0.06 &     0.11 &     0.11 \\ 
$\delta_{\mathrm{MSSM-weak}}[\%]$ &   2.73 &    -0.35 &    -0.35 \\ 
\hline
\end{tabular}
\end{center}
\mycaption{\label{Tab:results_SPS1b}  LO cross section in the
  improved Born approximation $\sigma_{\mathrm{IBA}}$ as defined in
  (\ref{eq:iba}), as well as  corrections $\delta$ relative to
  $\sigma_{\mathrm{IBA}}$, for  $\Pp\Pp\to
  (\Pb\bar{\Pb})\,\Phn/\PHn/\PAn\!+\!X$ at the LHC
  ($\sqrt{s}=14$~TeV) in the SPS~1b scenario. See
  Table~\ref{Tab:results_SPS4} for details.}
\end{table}
Because the masses of $\PAn/\PHn$ are larger and $\tanb=30$ is
smaller than in SPS~4, the total production cross sections are also
smaller. The cross sections including the cumulating effect of the
different higher-order corrections are shown in
Table~\ref{Tab:results_SPS1b_cross_sections}. 
\begin{table}
\begin{center}
\begin{tabular}{|l|c|c|c|c|c|c|}
\hline
& \multicolumn{2}{|c|}{$\Phn$} & 
\multicolumn{2}{|c|}{$\PHn$} &
\multicolumn{2}{|c|}{$\PAn$} \\ \cline{2-7}
&  $\Mb$[GeV]  & $\sigma$[pb] & 
 $\Mb$[GeV]  & $\sigma$[pb] & 
 $\Mb$[GeV]  & $\sigma$[pb] \\ \hline
QCD &                                 2.79 &     0.85 &     2.50 &     2.64 &     2.50 &     2.64 \\ 
+QED &                                2.79 &     0.84 &     2.50 &     2.63 &     2.50 &     2.63 \\ 
$+\Delta_{\Pb}^{\tilde{g}}$ &         2.76 &     0.83 &     2.08 &     1.82 &     2.08 &     1.82 \\ 
$+\Delta_{\Pb}^{\mathrm{weak}}$ &     2.77 &     0.83 &     2.30 &     2.24 &     2.30 &     2.24 \\ 
$+ \sin (\alpha_{eff})$ &             2.77 &     0.80 &     2.30 &     2.24 &     2.30 &     2.24 \\ \hline
full calculation &                    2.77 &     0.81 &     2.30 &     2.23 &     2.30 &     2.23 \\ 
\hline
\end{tabular}
\end{center}
\mycaption{\label{Tab:results_SPS1b_cross_sections} The NLO MSSM
  cross section $\Pp\Pp\to (\Pb\bar{\Pb})\,\Phn/\PHn/\PAn\!+\!X$ at
  the LHC ($\sqrt{s}=14$~TeV) in the SPS~1b scenario. See
  Table~\ref{Tab:results_SPS4_cross_sections} for details.}
\end{table}
The generic structure of $\delta_{\rm MSSM}$ as a function of $\MA$
with all other SPS~1b parameters fixed,
Fig.~\ref{Fig:MA_plot_SPS1b}, does not differ from SPS~4.

\end{appendix}

\end{document}